\documentclass[
  final            % use final for the camera ready runs
  ,numberedheadings % uncomment this option for numbered sections
  ]
  {aipproc}

\layoutstyle{8x11single}

\usepackage{amssymb}
\usepackage{axodraw}
\usepackage{graphicx}

\newcommand{\non}{\nonumber \\}

\newcommand{\alp}{\alpha}     \newcommand{\bet}{\beta}    
\newcommand{\gam}{\gamma}     \newcommand{\del}{\delta}   
\newcommand{\eps}{\epsilon}

\newcommand{\kap}{\kappa}     \newcommand{\lam}{\lambda} 
   \newcommand{\sig}{\sigma}   
  
\newcommand{\vphi}{\varphi}   \newcommand{\ome}{\omega}

\newcommand{\Gam}{\Gamma}     \newcommand{\Del}{\Delta}   
     \newcommand{\Lam}{\Lambda}

\newcommand{\cA}{{\cal A}}     
    \newcommand{\cD}{{\cal D}} 
    \newcommand{\cF}{{\cal F}}

    \newcommand{\cN}{{\cal N}} 
\newcommand{\cO}{{\cal O}}

\newcommand{\ZZ}{\mathbb{Z}}

\newcommand{\pa}{\partial} 
\renewcommand{\Box}{\square}

\newcommand{\inv}{^{-1}}

\newcommand{\rar}{\rightarrow}

\newcommand{\vac}{|0\rangle}

\newcommand{\xpv}[1]{\langle #1  \rangle}

\newcommand{\hlf}{\frac{1}{2}}
\newcommand{\ove}[1]{\frac{1}{#1}}

\newcommand{\bbbar}[1]{\overline{#1}}
\newcommand{\tttilde}[1]{\widetilde{#1}}

\newcommand{\vk}{\vec{k}}

\newcommand{\bpa}{\bet_{\parallel}}
\newcommand{\bpe}{\bet_{\perp}}
\newcommand{\bco}{\bet_{c}}

\begin{document}

\title{
The cosmological vacuum ambiguity, effective actions, and
transplanckian effects in inflation
}
\footnotetext{Presented by KS at the
  Mitchell Conference ``String Theory and Cosmology'', March 14-17,
  Texas A\&M.}

\author{Koenraad Schalm}{address={Institute for Strings, Cosmology and Astroparticle Physics,\\
Department of Physics, Columbia University, New York, NY
10027. \footnote{kschalm@phys.colombia.edu}
}
}

\author{Gary Shiu}{address={Department of Physics, University of
    Wisconsin, Madison, WI 53706. \footnote{
shiu@physics.wisc.edu}}
}

\author{Jan Pieter van der Schaar}{address={Department of Physics,
    CERN, Theory Division,  1211 Geneva 23. \footnote{jan.pieter.van.der.schaar@cern.ch}}
}

\begin{abstract}
We provide a prescription for parametrizing the vacuum choice
 ambiguity in cosmological settings. We introduce an arbitrary boundary action representing the initial conditions.
  A Lagrangian description is moreover the natural setting to study
 decoupling of high-energy physics. RG flow 
affects the boundary interactions. 
As a consequence 
 the boundary conditions are sensitive to high-energy physics through {\em irrelevant} 
 terms in the boundary action. Using scalar field theory as an example, we derive the 
 leading dimension four irrelevant boundary operators. We discuss how the known vacuum 
 choices, e.g. the Bunch-Davies vacuum, appear in the Lagrangian description and square 
 with decoupling. For all choices of boundary conditions encoded by relevant boundary 
 operators, of which the known ones are a subset, backreaction is under control. All, 
 moreover, will {\em generically} feel the influence of high-energy physics through 
 irrelevant (dimension four) boundary corrections. Having established a coherent 
 effective field theory framework including the vacuum choice ambiguity, we derive an 
 explicit expression for the power spectrum of inflationary density perturbations 
 including the leading high energy corrections. In accordance with the dimensionality 
 of the leading irrelevant operators, the effect of high energy physics is linearly 
 proportional to the Hubble radius $H$ and the scale of new physics
 $\ell = 1/M$. Effects of such strength are potentially observable in
 future measurements of the cosmic microwave background.
\end{abstract}

\maketitle

\section{Introduction}

The cosmological vacuum ambiguity has been a vexing problem for
decades now. In a spacetime background where the concept of energy
changes from observer to observer and time to time, we are still at a
loss how to unambiguously construct the quantum-mechanical ground state
--- or whether such a state even exists.\footnote{The notion of a
  ground state may be observer dependent.} 
For better or for worse, a consensus prescription has
emerged, the adiabatic/Bunch-Davies vacuum. Both solve a number of
conundrums, but leave others unanswered. A preference for either is
clearly a choice that is made. Initial conditions are always physical
input and rarely a consistency condition. 

The ambiguity shows in part why quantum field theory in a
curved, cosmological background is still an inexact science. We 
do not yet fully know how to quantize gravity. 
String theory does provide a fundamental framework to describe
gravitational physics at the highest energy scales. Yet, the details
of  
transplanckian physics, particularly in cosmological settings or how
they may affect vacuum selection, have completely eluded us so far. 
Fortunately, the notion of decoupling allows us to understand low
energy phenomena despite our ignorance of physics at very high energies. 
Renormalization Group (RG) flow teaches us that 
the effects of high energy physics can be captured by only a finite 
number of relevant couplings in the low energy theory. 
In flat spacetime, the decoupling between high and 
low energy physics is well established.
Again, however, for quantum field theories in curved space and in
FRW universes in particular, decoupling is not
so clearcut. In cosmological spacetimes 
high energy scales are redshifted to low energy 
scales via cosmic expansion. This connects 
high and low energy physics through unitary time evolution in addition
to the dynamics. 

Decoupling, specifically in the inflationary context, is of great
importance to 
upcoming cosmological precision experiments.
All current physical scales
would originate from transplanckian scales at the onset of inflation,
if inflation
lasted longer than the minimal number of e-folds. Conceivably,
then, signatures of 
Planck scale physics (stringy or other) could show up in cosmological measurements
\cite{Brandenberger,egks,ShiuWasserman,stanford,Others,Tanaka:2000jw,Lemoine:2001ar,Starobinsky:2002rp,Danielsson:2002kx,Burgess:2002ub}. 
This possibility
whether glimpses of transplanckian physics can be observed in the cosmic microwave
background (CMB) radiation \cite{Bennett:2003bz}  is determined by the
strength with which transplanckian physics decouples. 
Remarkably, such effects  {\em are} potentially observable, but only if 
the transplanckian physics selects a non-standard initial 
state \cite{egks,Danielsson:2002kx}.\footnote{In an abuse of language, we
  use vacuum and initial state interchangeably.} 
Other high energy effects are generically too small \cite{stanford} 
(with the exception of 
the higher dimensional operators identified in \cite{ShiuWasserman}).  
More recently, explicit examples were presented 
 to illustrate that the integrating out of a
massive field could result in a non-trivial initial state, 
offering both a proof of principle that transplanckian physics may be
observable, and suggesting that decoupling is more
subtle in expanding universes \cite{Burgess:2002ub}. 

\medskip 

In this review --- a condensation and expansion of \cite{nous} --- 
we would like to clarify the connections between
vacuum/initial state
selection and decoupling in a fixed FRW
background with the goal
to describe transplanckian effects in inflation (we ignore gravitational dynamics throughout). 
In cosmological settings, i.e. in a spatially homogeneous and 
isotropic universe, the size of the scale factor yields a preferred
time coordinate, and as a consequence a Hamiltonian approach 
has become standard \cite{BD}. In contrast to the Hamiltonian point of 
view which emphasizes the dynamical evolution, a Lagrangian point of
view emphasizes the symmetries and scaling behaviour
relevant to physical processes (see e.g. 
\cite{stanford,Burgess:2002ub,Larsen:2003pf}).  
It is therefore the natural framework for a
Wilsonian RG understanding of decoupling of energy scales and relevant
degrees of freedom determined by symmetries.\footnote{Wilsonian RG in
effect explains  why (non-gravitational) physics works. Its success strongly 
suggests that the same principles are at work in
{quantum gravity} and that general relativity is the low energy
effective action relevant at scales below $M_{Planck}$ (for a nice review on 
general relativity as an effective field theory see \cite{Burgess2}). String theory, in
particular, is an explicit manifestation of this idea.} 
However, a Lagrangian or an action by itself is insufficient to determine 
the full kinematic and dynamic behaviour of quantum fields. 
One must in addition specify the boundary conditions. 
This corresponds to the choice of initial or vacuum state in 
the Hamiltonian language. The question 
directly relevant to the window on transplanckian physics
provided by inflation is therefore which {\em boundary
  conditions} to impose on the fields. To preserve the symmetries of
the Lagrangian a subset of all possible boundary conditions is often
only allowed. With enough symmetry, e.g. Minkowski QFT, the choice may
in fact be unique.
FRW spacetimes have less symmetry and it is a priori not
clear, what the natural or correct boundary conditions
are. Here we re-encounter the cosmological vacuum ambiguity from the
Lagrangian perspective. How to proceed?

The clear advantage of the Lagrangian effective field theory formalism
is that at low energies the inital state will be determined by a
finite number of relevant boundary couplings. As always in effective
field theories, relevant couplings are determined by phenomenological
input: a measurement. The Lagrangian effective field theory formalism
therefore parametrizes our ignorance of the cosmological
initial conditions into measurable quantities. Clearly, this does not
solve the cosmological vacuum ambiguity, but it does give us a
quantitative controlled method to confront the ambiguity head-on. 
``When one does not know the answer, let a measurement decide''. 
 
What we will furthermore explain
in section \ref{sec:deco-theor-with} 
is that no matter which choice of boundary
conditions is made in the full quantum theory, RG-flow
in the effective low energy action will generically change these
conditions. In particular high-energy physics will affect
the boundary
conditions through irrelevant corrections, which we derive. 
We apply these results in section \ref{sec:transpl-effects-infl}
to the computation of the power spectrum of inflationary density
perturbations. The leading irrelevant correction to the boundary
conditions is of dimension four,
and we therefore find that the power spectrum is 
subject to corrections of order $H/M$ with $M$ the scale of new
physics. This is in accordance with
earlier predictions that transplanckian effects are potentially
observable \cite{egks,Danielsson:2002kx}. 
Importantly, we are able to derive this result purely
within the framework of Wilsonian effective field theory. This makes our answer
predictive both in the sense that the parametric dependence of
inflationary physics on high-energy is now manifest, and that the
strength is computable in any theory where the high energy physics is
explicitly known. Because our results are derived within 
the context of effective field theory, they provide a 
settlement to the debate
\cite{egks,stanford,Danielsson:2002kx,Brandenberger:2002hs} whether $H/M$ corrections are 
consistent with decoupling arguments.
We conclude with an outlook where we will briefly comment on the relation 
of our results to consistency issues regarding  
(non-trivial) de Sitter invariant vacua known as $\alp$-states.
We will, however, begin with a summary, lest the trees obscure the forest.

\subsection{The cosmological vacuum ambiguity, effective actions and
  transplanckian effects in inflation: a summary}
\label{sec:summary-our-results}

Any boundary conditions one wishes to impose can be encoded in a
boundary action. This is even true for the Minkowski vacuum (section
\ref{sec:mink-space-bound}). It has long been known 
that the couplings in  such a boundary
action are renormalized at the quantum level. Equivalently, a
Wilsonian approach to the effective action ought to result not only in
a renormalization of the boundary couplings, but also in the
generation of {irrelevant boundary} operators. Consider, for example, a two scalar field
model with a mass 
separation $M_{\chi} \gg m_{\phi}$ and boundary and
bulk interactions $S^{int} = - \int g\chi\phi -\oint
\gam\chi\phi$. This is exactly solvable, and upon integrating out
$\chi$, permitted when the cut-off scale $\Lam \ll M_{\chi}$, 
one generates the boundary interactions
\begin{eqnarray}
  \label{eq:34}
  S_{eff} = \oint \frac{g\gam}{M_{\chi}^2} \, \phi 
\frac{\Box^n}{M_\chi^{2n}}\phi~.
\end{eqnarray}
We will describe and review the Wilsonian effective action for
theories with a boundary, including this example, in
section \ref{sec:deco-theor-with}.

The issue of (boundary) Wilsonian decoupling is relevant to our
understanding of cosmology. In an expanding universe, there is no unique
vacuum state. In 
the Lagrangian language, this translates to a lack of
knowledge of the appropriate boundary conditions. Recall that {\em
  any} boundary conditions, including the `Minkowski' ones, can be
encoded 
in a boundary action. Wishing to emphasize the Lagrangian viewpoint,
where the 
study of decoupling is most natural, we add a boundary action with
free parameters at a fixed but arbitrary time $t_0$.

Our limited understanding of high-energy physics in the very early
universe can thus be accounted for by the
inclusion of a boundary action in a cosmological
effective Lagrangian. Whichever boundary conditions we choose this boundary
action to encode, they will be subject to
renormalization. In particular, 
the details of the high-energy physics, which has been
integrated out, will be encoded in irrelevant corrections to the
boundary action. 
For $\ZZ_2$ symmetric scalar field theory the leading irrelevant
boundary operators 
(that respect the homogeneity and isotropy of FRW cosmologies)
are
\begin{eqnarray}
  \label{eq:35}
  S_{bound}^{irr.op.} = \oint d^3x\left[
  -\frac{\bpa}{2M}\pa^i\phi\pa_i\phi - \frac{\bpe}{2M}
  \pa_{n}\phi\pa_{n}\phi
  -\frac{\bco}{2M}{\phi}\pa_n\pa_{n}\phi -\frac{\beta_4}{2M}\phi^4\right],
\end{eqnarray}
where $\pa_n$ is the normal derivative.
These operators are of dimension 
four --- one dimension higher than the boundary
measure ---  and describe corrections of order $|\vk|/M$ plus a
boundary four-point interaction. For the momentum range of interest to
the CMB, $|\vk| \sim H$, where $H$ is the Hubble parameter, the quadratic 
operators scale as $H/M$ and they are therefore the primary
candidates for witnessing consequences of high-energy physics in
cosmological data. The leading bulk operator is of order $H^2/M^2$
and is generically beyond observational
reach \cite{stanford}. Computing the inflationary perturbation
spectrum in a de Sitter background, including the corrections to 
Bunch-Davies boundary
conditions due to the irrelevant operators (\ref{eq:35}), we find 
\begin{eqnarray}
  \label{eq:40}
&& 
\hspace{-10pt} 
{P_{\!\!\!\!
\begin{array}{c}
\scriptscriptstyle 
BD +\\[-.12in]
\scriptscriptstyle 
irr.op.
\end{array}}^{\scriptscriptstyle dS}}
\!
= 
P_{\scriptscriptstyle BD}^{\scriptscriptstyle dS} 
\left( 1 -\frac{\pi }{4H} \left[\frac{\bbbar{H}_{\nu}^2(y_0)}{i}
\left[\frac{\vk_1^2(\bpa-\bco)}{a_0^2M} + \frac{\kap_{BD}^2\bpe}{M}
  -\frac{\bco m^2}{M} - \kap_{BD}\frac{3\bco H}{M}
\right] +{\rm c.c.}\right]
  \right), \non
\end{eqnarray}
with 
\begin{eqnarray}
\kap_{BD} &=& \frac{d-1+2\nu}{2} H -\frac{|\vk|}{a_0}
\frac{\bbbar{H}_{\nu+1}(y_0)}{\bbbar{H}_{\nu}(y_0)}~,
\end{eqnarray}
where $H_{\nu}(y_0)$ are Hankel functions at $y_0=|\vk|/a(\eta_0)H$
whose index $\nu(m^2)$ depends on the mass $m^2$. 
Crucial in our exposition
will be the proof (section \ref{sec:freed-choice-bound}) 
that, despite appearances, this expression does {not} depend on the location 
of the boundary action $y_0$.
Only the meaning of the initial conditions matters, 
not where they are imposed.

Eq. (\ref{eq:40}) is our main result. Having translated the cosmological 
vacuum choice ambiguity
into an arbitrary boundary effective 
action, we conclude based on Wilsonian
decoupling that the leading irrelevant operators in FRW field theory are 
boundary operators at order $H/M$. 
Using optimistic but not untypical estimates of 
$H \sim 10^{14}$ GeV  
and $M \sim 10^{16}$ GeV (string scale), new 
(transplanckian) physics will {\em generically} 
affect the standard predictions of inflationary cosmology
at the one-percent level. 
{\em Conversely, CMB observations with an accuracy of one percent or better
can potentially measure effects of transplanckian physics.}
Only for very special choices of
initial conditions and transplanckian physics will this correction be absent.

We further identify the boundary conditions corresponding to 
several cosmological vacuum choices including the generalization of
the ``Minkowski-space'' boundary conditions (sections
\ref{sec:mink-space-bound} and
\ref{sec:bound-cond-adiab}). In the Wilsonian effective Lagrangian
description it is clear that no vacuum is pre\-selected by a
consistency condition. Any boundary condition encoded by relevant
operators 
is consistent, in the
sense that the Minkowski space stress tensor counterterm generated
with the appropriate boundary conditions will
render the cosmological stress tensor finite as well (section
\ref{sec:pref-bound-cond}). Backreaction is always under
control. Which cosmological boundary conditions are the right ones to 
impose, requires just physical input, as it should be. 

\section{Decoupling in theories with a boundary: a review}
\setcounter{equation}{0}
\label{sec:deco-theor-with}

\subsection{Initial states in transition amplitudes, path integrals and
  fixed time\-slice boundaries}
\label{sec:init-stat-trans}

That boundary actions capture the initial conditions one
wishes to impose, follows directly from the relation
of the path-integral to quantum-mechanical transition amplitudes. We
will show this here. 

Recall that after a spatial Fourier transformation a field can be
considered as an infinite set of harmonic oscillators, each with
action
\begin{eqnarray}
  \label{eq:89}
  S^{bulk} = \int_{t_0}^{t_f}dt\left[ \frac{\dot{q}^2}{2}
  -\frac{\omega^2q^2}{2} \right]~.
\end{eqnarray}
This action is obtained from the quantum-mechanical transition
amplitude
\begin{eqnarray}
  \label{eq:90}
\int \cD x e^{iS^{bulk}} =  \langle x_N,t_f| e^{-i\hat{H}(t_f-t_0)}|x_1,t_0\rangle ~,~~~~~~
  \hat{H} 
  = \frac{\hat{p}^2}{2}+\ome^2\frac{\hat{x}^2}{2} ~,
\end{eqnarray}
by splitting the interval $t_f-t_0$ into $N$ smaller intervals of
length $(t_f-t_0)/N$, inserting $N-1$ complete sets of $|x\rangle$ and
$N$ complete sets of 
$|p\rangle$ states, and taking the continuum limit $N \rar \infty$.
This derivation makes clear that the action (\ref{eq:89}) has boundary
conditions $q(t_f)=x_N$, $q(t_0)=x_1$, and that the endpoints are {\em
  not} integrated over. Also clear is that {temporal} boundaries
are quantum-mechanically on a very different footing than spatial
boundaries. The latter simply affect the spatial
modefunctions. Temporal boundaries, however, are encoded in the choice
of initial and final state. In Lorentz-invariant field theory the
distinction disappears but for a technical point regarding reality
conditions that will become clear below.

For the free theory, a Gaussian integral, 
the exact answer for the transition amplitude is
easily obtained. One substitutes the solution to the field equation
with boundary conditions $q(t_f)=x_N$, $q(t_0)=x_1$ into the action.
Note that as the endpoints are not integrated over, 
the field equation is derived under the condition that the
variation $\del q$ vanishes on the boundary, $\del q(t_f)=0$, $\del
q(t_i)=0$. One finds the well-known results (up to normalizations,
which we ignore throughout this section)
\begin{eqnarray}
  \label{eq:91}
  q_{sol_1}(t) &=& De^{i\ome t} +{\rm c.c.}~~,~~~~D\equiv \frac{x_Ne^{-i\ome t_0}-x_1e^{-i\ome
  t_f}}{2i\sin(\ome(t_f-t_0))}\non
\int \cD x\,\, e^{iS^{bulk}} &=& \exp\left[-\ome\left(\frac{D^2(e^{2i\ome
  t_f}-e^{2i\ome t_0})}{2} -{\rm c.c.}\right)\right] \non
&\equiv& e^{iS^{bg,bulk}(x_N,x_1)} ~.
\end{eqnarray}

Consider now the transition amplitude for a different initial state. In
particular let us choose
the harmonic oscillator vacuum $|0\rangle$ annihilated by $\hat{a} =
\hlf\left(i\hat{p}+\ome\hat{x}\right)$. This corresponds to the
Minkowski space vacuum for the field mode with frequency $\ome$. The
transition amplitude $\langle x_N|e^{-i\hat{H}(t_f-t_0)}|0\rangle$
can be obtained from the standard transition amplitude by the
insertion of a complete set of states
\begin{eqnarray}
  \label{eq:92}
  \langle x_N|e^{-i\hat{H}(t_f-t_0)}|0\rangle = \int dx_1 \langle x_N|
  e^{-i\hat{H}(t_f-t_0)}|x_1\rangle\langle x_1|0\rangle~.
\end{eqnarray}
We can evaluate this expression in two ways. Either we can substitute
the harmonic oscillator ground state wave function $\langle
x_1|0\rangle \simeq e^{-\ome x_1^2/2}$ and the result (\ref{eq:91})
for the propagator. Performing the remaining Gaussian integral over
$x_1$, 
\begin{eqnarray}
  \label{eq:93}
  \int dx_1\, e^{iS^{bg,bulk}(x_N,x_1)}e^{-\frac{\ome x_1^2}{2}} =
  e^{-\frac{\ome x_N^2}{2}}~,
\end{eqnarray}
the result simply states that $|0\rangle$ is the zero energy
eigenstate of the (normal-ordered) Hamiltonian. (This is the usual way
one deals with non-trivial initial conditions in QFT.) Or we can again derive
a path-integral by
 splitting the interval $t_f-t_0$ into $N$ smaller intervals, now inserting $N$ complete sets of $|x\rangle$ and
$N$ complete sets of 
$|p\rangle$ states, and taking the continuum limit $N \rar
\infty$. Doing so yields the bulk action (\ref{eq:89}) plus a boundary
term
\begin{eqnarray}
  \label{eq:95}
  S^{bulk+bdy}= \int_{t_0}^{t_f}dt\left[ \frac{\dot{q}^2}{2}
  -\frac{\omega^2q^2}{2} \right] -\kap_0\frac{q(t_0)^2}{2}~,~~~\kap_0=-i\ome~.
\end{eqnarray}
The answer
for the transition amplitude $\langle
x_N|e^{-i\hat{H}(t_f-t_0)}|0\rangle$ ought then follow from solving
the field equations for this action including the boundary term, and
substituting the solution back. The extra insertion $\int dx_1
|x_1\rangle\langle x_1|$ means that the endpoint $q(t_0)$ is now
integrated over. The fluctuation $\del q(t_0)$ therefore no longer
vanishes and we obtain the field equations
\begin{eqnarray}
  \label{eq:96}
  \left(\frac{d^2}{dt^2}+\ome^2\right)q(t) = 0~,~~{\rm
  and}~~-\frac{d}{dt}q(t_0)-\kap_0q(t_0)=0~,
\end{eqnarray}
plus the implicit boundary condition $q(t_f)=x_N$.

We encounter a first subtlety. We wrote the action (\ref{eq:95})
in the
conventional way suggesting real boundary couplings. 
Yet the explicit computation shows that $\kap_0$ ought to be
imaginary. The subtlety lies in the reality condition for the
action. A check on the correct reality condition is that the Euclidean
path integral is damped. Clearly Wick rotating the boundary condition (\ref{eq:96})
compensates for the factor $i$, and all equations become manifestly real. The lesson is that the boundary
couplings for spacelike boundaries are imaginary. 
Still, because the coordinate $q(t)$ is manifestly real, one has to give a
prescription how to deal with the boundary condition (\ref{eq:96}) for
imaginary $\kap$. It is quite obvious that insisting on $q$ real,
i.e. $dq/dt(t_0)=0=q(t_0)$, or insisting that the action remain real,
$q^2 \rar |q|^2$, will not reproduce the known answer (\ref{eq:93}).
However, if we simply proceed on the assumption that $\kap$ is real,
i.e.
\begin{eqnarray}
  \label{eq:97}
  &&q_{sol_2}(t)= \cA(e^{i\ome t}+b_0e^{-i\ome t})~,\non
~~&& \cA(e^{i\ome
  t_f}+b_0e^{-i\ome t_f}) = x_N~,~~b_0 =
  -\frac{\kap_0+i\ome}{\kap_0-i\ome} e^{2i\ome t_0}~,
\end{eqnarray}
the answer for the background value of the action,
\begin{eqnarray}
  \label{eq:98}
  S^{bg,bulk+bdy} = \frac{i\ome}{2}\left(\frac{x_N^2}{(1+be^{-2i\ome
  t_f})^2} -\cA^2b^2e^{-2i\ome t_f} \right)~,
\end{eqnarray}
precisely reproduces the answer (\ref{eq:93}) 
for $\kap_0=-i\ome$ (hence $b=0$).
This is therefore the prescription for dealing with imaginary
boundary couplings: assume $\kappa$ is real until the final answer,
and only then analytically continue.

In the above example we have, of course, restricted ourselves to 
free field theory. One can repeat the whole exercise, however, with
the inclusion of a bulk source term $iS \rar iS+\int dt J(t)q(t)$
representing interactions. 
Treating the
source perturbatively, we expand into fluctuations $\xi$ around the
background solution, $q(t)=q_{sol}(t)+\xi(t)$. Integrating the
fluctuations out, we obtain for the action 
\begin{eqnarray}
  \label{eq:99}
  S_{\kap_f,\kap_0}^{bulk+bdy}(q) &=& \int dt \left[\frac{\dot{q}^2}{2}
  -\ome^2\frac{q^2}{2} -iJq\right]
  +\kap_f\frac{q(t_f)^2}{2}-\kap_0\frac{q(t_0)^2}{2}~,
\end{eqnarray}
the result
\begin{eqnarray}
  \label{eq:100}
  S^{bg,bulk+bdy}_{\kap_f,\kap_0}(J;q_{sol}) &=&
  S_{\kap_f,\kap_0}^{bg,bulk+bdy}(0) -i\int dt J(t) q_{sol}(t) \non
 && \hspace{1in}
-\frac{i}{2} \int dt dt'\,J(t)G_{\kap_f,\kap_0}(t,t')J(t')~.
\end{eqnarray}
where $G_{\kap_f,\kap_0}(t,t')$ is the Green's function obeying
$\frac{d}{d t} G(t,t')|_{t=t_0}= -\kap_0G(t_0,t')$, $\frac{d}{d t}
G(t,t')|_{t=t_f}= -\kap_fG(t_f,t')$ (see \cite{nous} for details). 
Note that at endpoints
where $q(t)$ is not integrated over, i.e. when $\del
q(t_{end})$ is constrained to vanish, $\xi(t_{end})$ also vanishes. At
these points the Green's functions for the fluctuations 
$\xi$ therefore obeys
Dirichlet boundary conditions with $\kap_{end}=\infty$. For the
transition function $\langle x_N|e^{-i\hat{H}(t_f-t_0)}|x_1\rangle$ we
thus have $\kap_f=\infty=\kap_0$, whereas for the transition function 
$\langle x_N|e^{-i\hat{H}(t_f-t_0)}|0\rangle$ we have $\kap_f=\infty$,
$\kap_0=-i\ome$. Equivalence between the two transistion functions
including bulk sources is thus established if
\begin{eqnarray}
  \label{eq:101}
\hspace{-15pt}
  \int \!dx_1
  \exp\left[{iS_{\kap_{f,0}=\infty}^{bg,bulk+bdy}(J;q_{sol_1}(x_N,x_1))}\right]
  \langle x_1|0\rangle
  = \exp\left[
{iS_{\begin{array}{c}
\scriptscriptstyle \kap_{f}=\infty, \\ [-.12in] \scriptscriptstyle \kap_0=-i\ome
\end{array}
}^{bg,bulk+bdy}(J;q_{sol_2}(x_N))}\right]\!\!.~
\end{eqnarray}
The only dependence on $x_1$ is in $q_{sol_1}(t)$ (eq. (\ref{eq:91})).
It is an instructive exercise to verify that eq. (\ref{eq:101}) is
indeed true. The prescription to deal with imaginary $\kap$ by  
analytic continuation to imaginary values in the final correlation functions,
therefore holds for perturbation theory as well.

This example is an explicit manifestion of the fact that (in
perturbation theory) all correlation functions are analytic in the
coupling constants. This necessarily includes boundary couplings,
which for a fixed time boundary correspond to initial conditions.

\subsection{Boundary field theory and RG flow}

The generalization from quantum-mechanical path-integrals to field
theory is straightforward. The difference of course
is that in field theory one has to address the infinities
encountered in the perturbation/loop expansion by renormalization. 
We review here, how
the boundary action is also affected by the renormalization procedure.
 
The study of field theories is primarily concerned with
Minkowski backgrounds, with the unique symmetry-compatible 
boundary conditions that the fields vanish at 
infinity.\footnote{One may alternatively
  think of Minkowski space field theory as defined on a (infinite
  volume) torus (``putting it in a box''), which has no boundary at
  all. See \cite{nous} for details.} 
Actions which contain explicit 
boundary interactions, however, have been studied in the past
\cite{Symanzik:1981wd,Candelas:qw,Deutsch:sc,Barton:1979yd}, and 
are receiving renewed attention (see e.g. 
\cite{Witten:2001ua,Fredenhagen:2003xf,Aharony:2003qf,Burgess1,Graham:2003nc,Jaffe:2003ji}).
As we have just shown, one can use such boundary 
interactions to enforce whichever boundary conditions one
wishes. Consider, for example, scalar $\lam\phi^4$ theory on a
semi-infinite space\footnote{We choose Lorentzian $+++-$ 
signature throughout 
the paper. Working with effective actions, we implicitly 
assume that all results can be obtained 
by a Wick rotation from Euclidean space. See the previous section
\ref{sec:init-stat-trans} for details on the Wick rotation.}
\begin{eqnarray}
  \label{eq:24}
  S_{bulk} = \int_{y_0 \leq y <\infty} \hspace{-0.4in}d^3 x dy\, - \hlf (\pa_{\mu} \phi)^2 -
  \frac{m^2}{2}\phi^2 -\frac{\lam}{4!}\phi^4~,
\end{eqnarray}
with the following boundary interactions added 
\begin{eqnarray}
  \label{eq:25}
  S_{boundary} = \oint d^3x - \frac{\mu}{2} \phi \pa_n \phi -
  \frac{\kap}{2}\phi^2 ~.
\end{eqnarray}
Here $\pa_n=\pa_y$ is the derivative normal to the boundary. Expanding the
action to first order in $\phi+\del\phi$, we find the
usual equation of motion 
\begin{eqnarray}
  \label{eq:26}
  \del S_{bulk} = \int d^3xdy \,\del\phi\left(\Box \phi - m^2\phi - \frac{\lam}{3!}\phi^3\right)~,
\end{eqnarray}
{\em plus} the boundary conditions
\begin{eqnarray}
  \label{eq:27}
  \del S_{bound} = \oint d^3x\, -\del\phi\left(\frac{\mu+2}{2}\pa_n\phi
  +\kap\phi\right)- \frac{\mu}{2} \phi\pa_n \del\phi ~.
\end{eqnarray}
If we insist that the variations $\del\phi$ are arbitrary and do not
vanish on the boundary (which would correspond to imposing Dirichlet 
boundary conditions), it appears that $\mu$ must vanish for consistency.
As we will see shortly, however, renormalization can
produce counterterms proportional to $\mu$ and a more correct point of
view is that $\phi$ can be discontinuously redefined on the
boundary \cite{Leigh:jq}, together with a redefinition of the
couplings 
which absorbs $\mu$:\footnote{
\label{fn:2}
Here $\theta(y)$ is the step function, with
  $\theta(0)=1/2$ and $\pa_y\theta(y)=\delta(y)$. 
  Recall that this distribution is of measure zero,
  i.e. \mbox{$\int_{y_0}^{\infty} dy \theta(y_0-y)f(y) =0$.} Of the bulk
  terms only the
  kinetic term is therefore affected by the shift. Also note
    that $\int_{y_0}^{\infty}\delta(y-y_0)f(y)=\hlf f(y_0)$.

One can also find a redefinition of the type
$\phi'(y)=\phi(y)+\alp\theta(y_0-y)\phi(y)$, which is the correct one 
from the point of view of coarse graining and the distributional
definitions for $\theta(y)$ and $\del(y)$. Interestingly, the
redefinitions required are the same.}
\begin{eqnarray}
  \label{eq:28}
   \phi(x,y) &\rar& \phi(x,y)+ \alp\theta(y_0-y)\phi(x,y_0)~, \non
   \kap' &\equiv & \kap+\kap\left(\alp+\frac{\alp^2}{4}\right)+
   \del(0)\left(\frac{\alp^2}{2}-\mu\alp-\frac{\mu\alp^2}{2}\right),
   ~~~~ \alp = \frac{2\mu}{(2-\mu)}.
\end{eqnarray}
This field redefinition can be interpreted as a 
shift of the boundary value of $\phi$ to the correct
saddlepoint.\footnote{When counterterms of the form $\phi\pa_n\phi$
  are required for renormalization, this shift of the background value
  for $\phi$ is thus a boundary analogue of the
  Coleman-Weinberg phenomenon.
} 
That this is the correct
interpretation follows from the fact that we can also treat $\mu$
perturbatively as an interaction. A Feynman diagram computation will
then yield an effective action with coupling $\kap'$.\footnote{A perturbative comparison
with Feynman diagrams explains the delta
function at zero argument \cite{nous}. 
It serves to make all distributions conform
to the bare boundary condition $\pa_n{\phi}= -\kap\phi$.} After 
this `renormalization' the boundary term from partial
integration is canonical 
\begin{eqnarray}
  \label{eq:29}
  \del S_{bound} = \oint d^3x -\del\phi\pa_n\phi - \kap' \del\phi \phi 
\end{eqnarray}
which vanishes when
\begin{eqnarray}
  \label{eq:30}
  \pa_n\phi = -\kap' \phi~.
\end{eqnarray}
We see that the (renormalized) value of $\kap$ determines the
boundary condition. 
For $\kap = 0$ we have Neumann boundary 
conditions, for $\kap =\pm \infty$ 
the (particular) 
Dirichlet boundary condition $\phi(x,y_0)=0$, 
and for finite $\kappa$ 
a mixture of the two. All 
possible (linear) boundary conditions are recovered.
This is comforting as there are 
no other terms of order $\phi^2$ 
compatible with the symmetries. In fact, the boundary
action $S_{bound}$ is the most general one we can write down,
if we limit our attention to {relevant} operators\footnote{
\label{fn:1b}
We assume that the initial state encoded by the boundary action
$S_{bound}$ has no intrinsic size, i.e. a 
dimensionful scale. We are ultimately interested in vacuum-like
initial conditions in cosmology. This restriction to scale-less
initial states is therefore a natural one.} 
and require (for
the sake of simplicity) that
the action is also invariant under the bulk $\ZZ_2$ symmetry $\phi
\leftrightarrow -\phi$. Of course, for a second order PDE one needs
two boundary conditions.
The other comes from the second boundary of
integration. In the example above this is $y=\infty$. See \cite{nous} 
for details.

RG arguments then 
tell us, that in a bounded space the terms in the boundary action, 
even if they 
were not present at
the outset, would be generated as
counterterms. They are
necessary for the consistency of the theory. Let us show this
explicitly. Suppose we start with Neumann boundary conditions: $\kap$
initially vanishes. By the method of images, the Neumann propagator
equals\footnote{Our domain of interest $y \in [y_0,\infty)$ is
  semi-infinite. Hence $k_y$ is a continuous variable.}
\begin{eqnarray}
  \label{eq:32}
  G_{N}(x_1,y_1;x_2,y_2) &=& -i\int \frac{d^3k_xdk_y}{(2\pi)^4}
\frac{e^{ik_x(x_1-x_2)}\left(e^{ik_y(y_1-y_2)}+e^{ik_y(-y_1-y_2+2y_0)}\right)} 
{k_x^2+k_y^2+m^2}~.
\end{eqnarray}
We will choose to regulate our theory by multiplying the propagator by
a regulating function $\cF(\Box/\Lam^2) = \exp(-k^2/\Lam^2)$
\cite{Polchinski}. 
This makes
the path integral well defined and cleanly separates out the
ultraviolet divergences.
The one-loop seagull graph then evaluates to
\begin{eqnarray}
  \label{eq:31}
&&\vspace*{-5in}
\SetScale{0.3}
\begin{picture}(300,30)(0,15)
\Line(50,0)(250,0)
\CArc(150,50)(50,0,360)
\Vertex(150,0){5}
\end{picture} \hspace{-2.5in} =
\langle \phi(x_1,y_1)\phi(x_2,y_2)\rangle_{1-loop} \nonumber \\[.3in]
&&\hspace{0.2in}=\hspace{0.2in}\frac{-i\lam}{4}
   G_N(x_1,y_1;x_1,y_1)\del^3(x_1-x_2)\del(y_1-y_2) \non
&&\hspace{0.2in}=\hspace{0.2in}
\frac{-\lam\del^3_{x;1,2}\del_{y;1,2}}{4(2\pi)^4} 
\left[ \int d^4k \frac{e^{-\frac{k^2}{\Lam^2}}}{k^2+m^2}+  
       \int d^3k_xdk_y \frac{e^{ik_y(-2y+2y_0)-\frac{k^2}{\Lam^2}}}
                      {k_x^2+k_y^2+m^2}
\right] ~.
\end{eqnarray}
The first term is the usual bulk $\lam\phi^4$ divergence 
of the two-point function. 
The second
term, however, is a newly divergent term, and quite obviously a direct
consequence of the boundary conditions. Evaluating this term in more
detail, we find 
\begin{eqnarray}
 \label{eq:50}
\xpv{\phi\phi}_{1-loop} &=&
    \frac{\lam\del^3_{x;1,2}\del_{y;1,2} }{4(2\pi)^4}
    \left(
      \pi^{5/2} \Lam e^{\frac{m^2}{\Lam^2}}
    \right) 
    \left(
      \frac{\Lam}{\sqrt{\pi}} \int_{0}^{1} ds
      e^{-s\Lam^2(y_0-y)^2-\frac{m^2}{\Lam^2s}} 
    \right) 
    \non
&\sim&\left.
  \lam \Lam \del_x^3\del(y_1-y_2)\del(y_1-y_0)\right|_{\Lam^2 \gg m^2} ~.
\end{eqnarray}
Note that the new divergence is entirely located on the boundary. 
The last step utilizes one of the more common distributional
definitions of the Dirac-delta function (before doing the finite
integral over $s$). Recalling the coarse-graining
steps underlying RG-flow, it should come as no surprise that
the delta-function localization appears in a distributional
limit. This simply reflects that our spatial resolution decreases
under RG-flow, and the precise location of the boundary becomes fuzzy.

That the divergence is concentrated solely on the boundary (in this
distributional sense)
is reassuring. Bulk
UV-physics should be unaffected by the presence of a boundary. It is
precisely the breaking of Lorentz invariance due to
the presence of the boundary that
is responsible for the new divergence. By necessity it must then
appear in the same sector of the theory that was responsible for the
symmetry-violation in the first place. 

To make the theory finite, we therefore need to add a boundary
counterterm of the type\footnote{Since 
  the `bare' boundary conditions are Neumann, this is the only type we
  can add.}
\begin{eqnarray}
  \label{eq:46}
  S_{bound}^{count} = \oint_{y=y_0} d^3x \,\,
   \xi^2(m^2/\Lam^{2})\, \left(
    \frac{\lam\Lam}
         {\pi^{3/2}}
       \right) \phi^2~.
\end{eqnarray}
with $\xi(m^2/\Lam^{2})$ 
chosen such that it cancels the divergence in eq. (\ref{eq:50}). 
This result is of course expected (in part) purely on dimensional grounds.

The necessity of this counterterm has serious implications, however. 
Recalling the results from the first half of this section, we see that
the boundary conditions {\em change} under RG-flow. In order to
reproduce the same physics in a theory with a different cut-off, we
not only need to change the vertices, but also the {\em boundary
  conditions}. (More precisely, to maintain a given physical renormalized
boundary condition $\kap_{ren}$ we need to change the bare coupling $\kap$.)
Of course, this counterterm is
scheme-dependent.
The beta-functions at one loop on the other hand are
scheme-independent, and we can extract the generic behaviour of the
boundary conditions from them. 
We find that as we change the scale, the boundary
conditions {change} under RG-flow as
\begin{eqnarray}
  \label{eq:33}
  \beta_{\kap} &\equiv&\left. \Lam \frac{\pa \kap}{\pa
   \Lam}\right|_{m^2/\Lam^2 
   {\rm fixed}} \hspace{-.025in}= \xi^2 \Lam
  \frac{\lam}{\pi^{3/2}} + \cO(\lam^2) ~.
\end{eqnarray}
with $\xi^2 >0$.
This may seem surprising, but it does {not} go against
the lore that boundary conditions are determined by physical
conditions, and not by dynamics. It is worthwhile to repeat 
that what the
RG-scaling of the boundary conditions says, is that in a {\em cut-off}
theory, under a change of the cut-off, one reproduces the same physics
when one changes the boundary 
conditions according to eq. (\ref{eq:33}). 

\subsection{Boundary RG fixed points and `vacua'}

A natural question to ask is what the endpoints of boundary
RG-flow are.
The explicit dimensionality of the coupling $\kap$ already betrays the
answer. In the deep IR, when $|p| \ll \Lam$ ($\Lam \rar \infty$
effectively; $m =\mu\Lam$), $\kap$ blows up, and the
boundary conditions tend to the special Dirichlet boundary condition 
$\phi(x,y_0)=0$. Physically this is easily understood 
in 
Wilsonian RG
language. The moment the cut-off restricts the momentum scales $|p|$
to be smaller than $m$ ($\Lam \sim m$), all modes freeze out and the theory 
ceases to be dynamical. Hence the field $\phi$ `vanishes', and must be
Dirichlet. 

Dirichlet conditions
thus form a trivial
fixed point of RG-flow.
This is easily visible.  
When $\phi$ strictly vanishes on the boundary, simply no
counterterms are possible. Both terms $\oint \phi\pa_n\phi$ and $\oint
  \phi^2$ vanish. 
For completeness, were one to repeat the computation
eq. (\ref{eq:31})
for Dirichlet conditions, the difference is that the
propagator now has a relative minus sign. As a consequence, 
the bulk divergence {cancels} the boundary divergence at $y=y_0$. Eq. (\ref{eq:31}) shows
  this clearly. 
In effective field theory the distinction between the fuzzy 
boundary and the bulk disappears in the deep IR 
limit, which explains why we can no longer 
treat bulk and boundary singularities separately 
when the boundary conditions become Dirichlet.

When the boundary is spacelike and represents initial conditions in time,
the induced changes in the boundary conditions due to RG-flow have a
natural description in the Hamiltonian language of states. Under
coarse graining the original state gets screened by vacuum
polarization. In the low-energy effective theory, the correct state
to use is a dressed version of the original state. If we take this
picture further, we can deduce the boundary conditions which
correspond to the vacuum. If the vacuum is the `empty' state, then it
ought not to become dressed under coarse graining. Translating back
to the Lagrangian language, this means that the corresponding boundary
conditions will not suffer from renormalization. Hence a vacuum in
the Hamiltonian language should correspond to a fixed point of boundary
RG-flow.\footnote{Presumably this is a 
UV-fixed point. Exciting the vacuum to a state, i.e. deforming away 
from the fixed point, reinstates RG-flow. 
The excitation, however, should not disappear in the deep IR. 
Hence the dressing of the state due to coarse graining
leads one away from the vacuum. Of course to study boundary RG-flow,
one needs an interacting theory. Any state in a  
free theory is a trivial fixed point of boundary RG-flow.}

\subsection{Freedom of choice for the boundary location}
\label{sec:freed-choice-bound}

What will be of fundamental importance to us, is that the location of
the boundary is arbitrary. The introduction of a boundary action at
$y_0$ is a way to encode the initial conditions at the level of the
action, but it does {not} necessarily mean that there is a
physical object or obstruction at $y=y_0$. It is simply a translation
of the statement that a second order PDE needs two boundary conditions,
but at what location one imposes those conditions is irrelevant. Of
course, if one imposes the boundary conditions at a different location,
they will not in general  be of the same form as the original initial
conditions. If one changes the location $y_0$ one must change the
value of $\kap$ to keep the physics unchanged. A {symmetry} is
therefore present between the location $y_0$ and $\kap$.\footnote{This
  is not a true symmetry of the action. Because the coupling constant
  $\kap$ changes, it is an isomorphism between families of
  theories. This is analogous to general coordinate invariance of the
  target space manifold in non-linear sigma models.}
 To show this
explicitly, choose a basis $\vphi_+(\vk,y)$, $\vphi_-(\vk,y)=\vphi_+^{\ast}(\vk,y)$ 
for the two independent solutions of the kinetic operator. In terms of
this basis, the linear combination
which obeys the boundary condition $\pa_n\vphi(y_0)=-\kap\vphi(y_0)$ is
\begin{eqnarray}
  \label{eq:15}
\vphi_{b_{\kap}}(\vk,y) \equiv  \vphi_+(\vk,y)+b_{\kap}(\vk)\vphi_-(\vk,y)~,~~~ b_{\kap}(\vk) =
  -\frac{\kap\vphi_{+,0}+\pa_n\vphi_{+,0}}{\kap\vphi_{-,0}+\pa_n \vphi_{-,0}}~,
\end{eqnarray}
Here the subscript $0$ means that the quantity is 
evaluated at the boundary $y_0$.
Obviously if $b_{\kap}$ stays the same, physics stays the same. This allows
us to derive a symmetry relation between the value $\kap$ and the
location $y_0$. Under a constant shift of the boundary 
$\del\vphi = \xi\pa_n\vphi = \xi\pa_y\vphi$ and a simultaneous change $\del
\kap$, $b_{\kap}$ 
changes as\footnote{
Note that $b_{\kap}$ depends on the basis choice $\vphi_{\pm}$, but
$\kap$ does not.}
\begin{eqnarray}
  \label{eq:14}
  \del b_{\kap} &=&
  -\xi\left[
\frac{\kap\pa_n\vphi_{+,0}+\pa_n^2\vphi_{+,0}}{\kap\vphi_{-,0}+\pa\vphi_{-,0}} 
    -
  \frac{\kap\vphi_{+,0}+\pa_n\vphi_{+,0}}{(\kap\vphi_{-,0}+\pa_n\vphi_{-,0})^2}
  (\kap\pa_n\vphi_{-,0}+\pa_n^2\vphi_{-,0})\right]
  \non
&&-\del\kap\left[
  \frac{\vphi_{+,0}}{\kap\vphi_{-,0}+\pa_n\vphi_{-,0}}
  -
  \frac{\kap\vphi_{+,0}+\pa_n\vphi_{+,0}}{(\kap\vphi_{-,0}+\pa_n\vphi_{-,0})^2}
  (\vphi_{-,0})\right]~. 
\end{eqnarray}
Demanding that $\del b_{\kap}$ vanishes, one finds the change in $\kap$
necessary to keep physics unchanged under a change of the location of
the boundary.  This shows explicitly that this location is {arbitrary}.

\subsection{Minkowski space boundary conditions}
\label{sec:mink-space-bound}

Minkowski space formally does not have a boundary of course. The
{arbitrariness} of the location of the boundary, however,
suggests that we should be able to treat it in a similar way. 
This is not quite manifest because, to stay within the framework of effective 
field theory, $\kap$ must remain an analytic dimension
one operator in the spatial momenta. The symmetry (\ref{eq:14}) is
subject to this condition. The harmonic oscillator boundary conditions, constructed
here to yield physics equivalent to unbounded 
Minkowski space physics, will be consistent with this requirement.
To find these conditions suppose the boundary is a
fixed time slice. We can then take a 
cue from the
Hamiltonian formalism. Minkowski boundary conditions should 
correspond to choosing the standard 
Minkowski vacuum in the Hamiltonian
picture. By definition this is the state 
annihilated by the lowering operator of each {spatial} momentum
mode $\vk_x$ (in the free theory).
\begin{eqnarray}
  \label{eq:59}
  \hat{a}_{\vk} \vac =0 ~~~\Leftrightarrow ~~~\left( \hat{\pi}_{\vk}
-i\ome(\vk,m) \hat{\phi}_{\vk} \right) \vac =0~,~~~~~ \ome(\vk,m) = 
\sqrt{\vk^2+m^2}~.
\end{eqnarray}
The canonical momentum conjugate to $\pi_k=\pa_0\phi_k$ is precisely the
normal derivative to the fixed time slice. 
This suggests that we should choose the {spatial} 
momentum dependent boundary \mbox{conditions
\cite{Hamilton:2003xr}}
\begin{eqnarray}
  \label{eq:86}
  \pa_n \phi|_{y=y_0} = i\sqrt{\vk^2+m^2}\, \phi|_{y=y_0}   
  ~~~\longrightarrow~~~ 
  \kap= -i\sqrt{\vk^2+m^2}~.
\end{eqnarray}
This boundary condition descends from the 
`higher derivative' operator $\oint \phi \sqrt{\pa_i^2-m^2}\phi$. But, as $\kap$ has canonical dimension one, there is no new
scale associated with this higher derivative term. Note that $\kap$
is purely imaginary. We recall from section \ref{sec:init-stat-trans}
that this is a consequence of imposing the boundary
condition at a fixed time. Wick rotating from a spatial boundary with
real $\kap$ generates a factor of $i$ in the boundary condition
$\pa\phi=-\kap\phi$. All correlation functions will be analytic in
the boundary coupling $\kap$, as is usual in
effective field theory, and we are therefore instructed to treat
$\kap$ as real throughout all steps of the calculation, substituting
its imaginary value only at the end.

This momentum dependent choice of boundary conditions
indeed ensures that the theory reproduces Minkowski space dynamics. For
an arbitrary $\kap$ the Green's function is (see eq. (\ref{eq:15}),
and recall that $y$ parametrizes a timelike direction) 
\begin{eqnarray}
  \label{eq:57}
  G_{\kap}(x_1,y_1;x_2,y_2) = -i\int \frac{d^3\vk dk_y}{(2\pi)^4}
  \frac{e^{i\vk(x_1-x_2)}\left(e^{ik_y(y_1-y_2)}+\frac{ik_y+\kap}{ik_y-\kap}e^{ik(-y_1-y_2+2y_0)}\right)}{\vk^2-k_y^2+m^2
  -i\eps} ~,~~
\end{eqnarray} 
where we have included the $i\eps$ term. The second term, at first
sight, negates equivalence with the Minkowski propagator
\begin{eqnarray}
  \label{eq:56}
  G_{Mink} = -i\int \frac{d^3\vk dk_y}{(2\pi)^4} \frac{e^{i\vk(x_1-x_2)+ik_y(y_1-y_2)}}{\vk^2-k_y^2+m^2-i\eps}~,
\end{eqnarray}
The coefficient $\kap$, however, is precisely chosen such that {on
  shell} the second term vanishes.\footnote{The second term only
  vanishes for 
the domain $\theta(y_1+y_2-2y_0)$. Since our domain of
  interest is $y>y_0$, this is always true.} 
By unitarity, the
  theory with $\kap = -i\ome(\vk,m)$ is then the same as the
  Minkowski space theory. We can see this explicitly by performing the
  integral over $k_y$. Doing so returns the standard Minkowski
  propagator in Hamiltonian form 
\begin{eqnarray}
    \label{eq:87}
     G(x_1,y_1;x_2,y_2) = \int \frac{d^3k}{(2\pi)^3}\frac{
     e^{i\vk(\vec{x}_1-\vec{x}_2)-i\ome(\vk,m)(y_1-y_2)}}{2\ome} \theta(y_1-y_2) +
     (y_2 \leftrightarrow y_1)~,
\end{eqnarray}
which shows that the second term really is spurious. Indeed, this
choice of $\kap$ removes the pole in the second term, which means its
contribution to any physical quantity disappears. 

We still have an official boundary at $y_0$ of course, even though the
specific boundary conditions (\ref{eq:86}) ensure that it has no
effect on physical amplitudes. 
The situation described here, is familiar from 
electrodynamics.\footnote{Except that this boundary
is spacelike, which is why we can in fact relate it to a choice of
initial state.}
We have chosen an interface at $y_0$ where the
dielectric properties happen to be the same for both materials. The
transmission coefficient 
is therefore 100\% and the wavefunction behaves as if the interface is
not there, i.e. the interface is completely transparent.

\subsubsection{Minkowski boundary conditions and RG-flow}

Classical physics is indeed insensitive to a completely transparent
interface. Is the quantum physics as well? In other words does the fact
that the off-shell
propagators appear to differ become relevant at the loop level? The
answer is obviously no in perturbation theory. The cancellation of the pole by the specific
`Minkowski' choice for $\kap$ means that in any integral the
contribution of the second term vanishes. Hence the Minkowski boundary
conditions do not get renormalized. They are a fixed point of boundary
RG-flow exactly as befits the boundary conditions corresponding to a
true vacuum. 
The reason why this is so is clear. The choice
$\kap_{Mink} = -i\ome(\vk,m)$ is precisely the one that restores the
Lorentz symmetry naively broken by the introduction of a
boundary. Counterterms are forbidden to appear for they would break
the reinstated Lorentz symmetry.

\subsection{Wilsonian RG-flow and irrelevant operators}
\label{sec:wilsonian-rg-flow}

Quite generically 
therefore the boundary conditions of a 
quantum field theory are affected by RG flow, unless they are protected 
by a symmetry. Integrating out
high energy degrees of freedom necessitates a change in boundary
conditions 
to reproduce the same physics in a low-energy 
effective description of the
theory. Decoupling then ensures that the low-energy theory remains
predictive: the effects of high-energy physics are primarily encoded in a
small set of relevant operators with universal scaling behaviour
independent of the details of the high-energy theory. 
Subleading corrections 
of an energy expansion 
are by definition captured by irrelevant
operators. These encode the specifics of the high-energy completion of
the theory. 

One of
our best hopes to detect the properties of high energy
physics beyond the Planck scale is in a
cosmological setting. The tremendous cosmological redshift during 
inflation may bring the consequences of such irrelevant
operators within reach of experimental measurements. This exciting
opportunity has been a preeminent question in recent literature. In
section \ref{sec:transpl-effects-infl} we shall show that the
irrelevant {boundary} operators discussed in this subsection are
responsible for the {leading} effects of high-energy physics in
cosmology, appearing generically at order $H/M_{Planck}$.
The leading irrelevant operators for the bulk theory have long been
known and  their consequences for cosmological measurements
are discussed in \cite{stanford}. However, it is well known 
that quantum field theory in cosmological settings suffers
from a vacuum choice ambiguity. In the Lagrangian language this
corresponds to a choice of boundary conditions. As we have just seen, we can 
parametrize this ambiguity 
in the cosmological vacuum choice by adding an
arbitrary boundary action $\oint\kap\phi^2$. 
Whichever the value of
$\kap$ may be, the influence of high-energy physics
will be encoded in the irrelevant corrections to the boundary action. 
For that
reason, we devote this section to a determination of  
the leading irrelevant operators 
on the boundary. Earlier studies have 
indeed indicated it is 
only  (irrelevant) changes in the
boundary condition which can have observable effects in
measurements. Due to the symmetry constraints on the action the
consequences of bulk irrelevant operators are just too small to be
detectable. Our aim here is to provide a solid foundation for these
earlier results.

One can make a straightforward guess as to 
what the leading boundary irrelevant operators are, insisting on
locality, compatibility with the $\ZZ_2$ symmetry, and $SO(3)$ rotational
invariance on the boundary.\footnote{
These symmetry constraints follow from the assumption that the initial
state has no intrinsic dimensionful parameter.
See footnote \ref{fn:1b}.}
They are the
dimension four operators: 
\begin{eqnarray}
  \label{eq:42a}
  \oint_{y=y_0}\hspace{-.2in}d^3x\,\, \phi^4~,~~~
  \oint_{y=y_0}\hspace{-.2in}d^3x\,\, \pa^{i}\phi\pa_{i}\phi~,~~~ 
  \oint_{y=y_0}\hspace{-.2in}d^3x\,\, \pa_n\phi\pa_n\phi~,~~~ 
  \oint_{y=y_0}\hspace{-.2in}d^3x\,\, \phi\pa_n\pa_n\phi~.
\end{eqnarray}
Note that the breaking of Lorentz invariance on the boundary
distinguishes normal and tangential derivatives, and that normal
derivatives cannot be integrated by parts. Varying $\phi$
infinitesimally, the latter two will generate normal derivatives on the
variation $\pa_n\del\phi$. To restore the applicability of the
calculus of variations, one needs to perform a discontinuous field redefinition and
adjustment of the couplings similar to (\ref{eq:28}). (For the
interested reader, we do so in \cite{nous}.) 
In this sense, all physics can be captured by 
the first two irrelevant
operators. 
However, for tractability we will treat all four operators
perturbatively and on the same footing. 
We will see in section \ref{sec:transpl-effects-infl} that these
operators will lead to corrections of order $H/M_{Planck}$ 
to inflationary density perturbations, as 
predicted by the studies \cite{egks}. 
Here we will give an explicit example where high-energy
physics induces two of these
dimension four irrelevant boundary operators.

Tree-level diagrams exchanging a heavy field are the natural
candidates for producing higher derivative corrections under RG-flow.  
We therefore add a scalar $\chi$ to the theory with mass $M_{\chi} \geq
\Lam$, to represent the high energy sector whose influence we will
deduce. The only communication between the field $\chi$ and $\phi$ will
be through the `flavor-mixing' bulk and boundary couplings 
\begin{eqnarray}
  \label{eq:41}
  S_{high}^{int} = - \int d^3xdy\,\, g\chi\phi~ - \oint d^3x \,\,\gam\chi\phi~,
\end{eqnarray}
and $\chi$ will have no other bulk or boundary
(self)-interactions. 
Because the mass of $\chi$ is higher than the
cut-off, it will not appear as a final state, and in this simple model
we can integrate it
out explicitly. Its influence on the low-energy effective $\lam\phi^4$
theory is only through  tree-level mass oscillation graphs and a boundary
reflection. Treating the couplings $g$ and $\gam$ as perturbations ---
hence the propagator for $\chi$ will have Neumann boundary 
conditions ---
consider the tree level correction to
$\xpv{\phi\phi}$ represented by the following Feynman diagram
and its effective replacement.
\begin{eqnarray}
  \label{eq:47}
    \SetScale{0.8}
     \begin{picture}(300,50)(100,5)
   \Line(225,45)(200,65)
   \Photon(225,45)(250,25){5}{5}
   \Line(250,25)(300,45)
   \SetColor{Gray}
   \Line(250,27)(260,23)
   \Line(260,27)(270,23)
   \Line(270,27)(280,23)
   \Line(280,27)(290,23)
   \Line(290,27)(300,23)
   \Line(300,27)(310,23)
   \Line(310,27)(320,23)
   \Line(170,27)(180,23)
   \Line(180,27)(190,23)
   \Line(190,27)(200,23)
   \Line(200,27)(210,23)
   \Line(210,27)(220,23)
   \Line(220,27)(230,23)
   \Line(230,27)(240,23)
   \Line(240,27)(250,23)
%   \GBox(170,23)(330,27){.9}
   \SetColor{Black}
   \DashLine(250,25)(250,0){5}
%Arrow
   \Text(290,25)[r]{$\Longrightarrow$}
%  Second Graph
   \Line(400,45)(450,25)
   \Line(450,25)(500,45)
   \SetColor{Gray}
   \Line(450,27)(460,23)
   \Line(460,27)(470,23)
   \Line(470,27)(480,23)
   \Line(480,27)(490,23)
   \Line(490,27)(500,23)
   \Line(500,27)(510,23)
   \Line(510,27)(520,23)
   \Line(370,27)(380,23)
   \Line(380,27)(390,23)
   \Line(390,27)(400,23)
   \Line(400,27)(410,23)
   \Line(410,27)(420,23)
   \Line(420,27)(430,23)
   \Line(430,27)(440,23)
   \Line(440,27)(450,23)
%   \GBox(170,23)(330,27){.9}
   \SetColor{Black}
   \DashLine(450,25)(450,0){5}
  \end{picture}
\end{eqnarray}
Here wiggled lines denote the heavy field $\chi$, solid lines the
light field $\phi$; the shaded region denotes the boundary, and the
dashed line the insertion of a $\gam$-vertex. This diagram is easily
evaluated to
\begin{eqnarray}
  \label{eq:62}
  \langle \phi(x_1,y_1)\phi(x_2,y_2)\rangle_{\chi-effect} 
  &=& -2 g \gamma
    G_N(x_1,y_1;x_2,y_0)\del(y_2-y_0) \non
  &=&
    \frac{2ig\gam \del(y_2-y_0)}{(2\pi)^4} 
    \left[ \int d^4k
  \frac{e^{ik_x(x_1-x_2)+ik_y(y_1-y_0)-\frac{k^2}{\Lam^2}}}{k^2+M_{\chi}^2} 
    \right] ~.~~~~~~
\end{eqnarray}
Approximating the denominator in the standard way by a geometric series
valid for \mbox{$M_{\chi}^2 \gg \Lam^2$},
\begin{eqnarray}
  \label{eq:48}
    \xpv{\phi\phi}_{\chi} &=&
    \frac{2ig\gam\del_{y_2-y_0}}{M_{\chi}^2(2\pi)^4} \sum_{n=0}^{\infty}
    \left[ \int d^4k \left(\frac{-k^2}{M^2_{\chi}}\right)^n 
      e^{ik_x(x_1-x_2)+ik_y(y_1-y_0)-\frac{k^2}{\Lam^2}}
    \right] ~,
\end{eqnarray}
we extract the $k_y$ dependence in the second term as a derivative to
find\footnote{Note that these results are not inconsistent
    with our earlier 
calculation (\ref{eq:50}). 
There we evaluate the answer 
in the approximation
 $\Lam \gg m$. Here we approximate $\Lam
    \ll M_{\chi}$. The exact intermediate answer obtained in
  eq. (\ref{eq:50}) is non-perturbative in $\Lam/M$. This is why we 
  approximate the momentum integral for $M_{\chi} \gg \Lam$ 
in the standard way.}
\begin{eqnarray}
  \label{eq:60}
  \xpv{\phi\phi}_{\chi} &=&
    \frac{2ig\gam\del_{y_2-y_0}}{M_{\chi}^2(2\pi)^4} \sum_{n=0}^{\infty}
    \left[ \left(\frac{\Box_1}{M^2_{\chi}}\right)^n \int d^4k
      e^{ik_x(x_1-x_2)+ik_y(y_1-y_0)-\frac{k^2}{\Lam^2}}
    \right] \non
  &=&
    \frac{2ig\gam\del_{y_2-y_0}}{M_{\chi}^2(2\pi)^4} \sum_{n=0}^{\infty}
    \left[ \left(\frac{\Box_1}{M^2_{\chi}}\right)^n \Lam^4\pi^2e^{-\Lam^2\frac{(x_1-x_2)^2}{4}-\Lam^2\frac{(y_1-y_0)^2}{4}}
    \right] ~.
\end{eqnarray}
Now recall that the
projection onto the boundary of bulk terms appears as a 
distribution with resolution $\Lam$. In this sense the
above term contains the delta function $\frac{\Lam}{2\sqrt{\pi}}
e^{-\Lam^2 (y-y_0)^2 /4}$. 
Up to this resolution the above expression is thus 
equivalent to 
\begin{eqnarray}
  \label{eq:61}
  \xpv{\phi\phi}_{\chi} 
  &=& \frac{2ig\gam\del_{y_2-y_0}}{M_{\chi}^2} \sum_{n=0}^{\infty}
    \left[ \left(\frac{\Box_1}{M^2_{\chi}}\right)^n \del_{\Lam}^3(x_1-x_2)\del_{\Lam}(y_1-y_0)    \right] ~.
\end{eqnarray}
Hence we see explicitly the resultant higher derivative {boundary}
interactions
in the $\phi$ low-energy effective action. The above results
correspond to the vertices
\begin{eqnarray}
  \label{eq:63}
  S_{eff} =  \oint d^3x \frac{g\gam}{M_\chi^{4}} \left[
  \pa_{i}\phi\pa^{i}\phi -\phi\pa_n\pa_n\phi\right] + \cO((\pa/M)^4) ~.
\end{eqnarray}
This supports the naive integrating out of $\chi$ after a shift $\chi
\rar \chi - g(\Box+M^2)\inv \phi$ as argued in section
\ref{sec:summary-our-results}. The terms arising from the boundary term
$\oint \gam\chi\phi$ under this shift precisely reproduce the higher 
derivative terms (\ref{eq:63}).

Note the similarity between the expression (\ref{eq:62}) and the
image-charge term in the seagull-graph (\ref{eq:31}). We see therefore
that a similar set of higher derivative corrections can arise from
{\em loop}-diagrams in a $\chi\phi$ theory with only the bulk
interaction
\begin{eqnarray}
  \label{eq:64}
  S_{high}^{int} = \int d^3xdy -\tttilde{g}\chi^2\phi^2~.
\end{eqnarray}
This is the hybrid inflation inspired model, considered before in the
context of decoupling in FRW-spacetimes \cite{Burgess:2002ub}.
The seagull diagram responsible for the  higher-derivative corrections
is a direct copy of eq. (\ref{eq:31}) only to be
evaluated in the limit $M_{\chi} \gg \Lam$ rather than $m_{\phi}\ll \Lam$.
\begin{eqnarray}
\label{eq:62b}
&&\vspace*{-5in}
\SetScale{0.3}
\begin{picture}(300,30)(0,15)
\Line(50,-25)(150,0)
\Line(150,0)(250,-25)
\PhotonArc(150,50)(50,-90,270){-6}{8.5}
\Vertex(150,0){5}
\end{picture} \hspace{-2.5in} =
\langle \phi(x_1,y_1)\phi(x_2,y_2)\rangle_{\chi-effect} \nonumber
\\[.3in]
  &&\hspace{0.2in}=\hspace{0.2in} -i\tttilde{g}
    G_N(x_1,y_1;x_1,y_1)\del^3(x_1-x_2)\del(y_1-y_2) \non
  &&\hspace{0.2in}=\hspace{0.2in}
    \frac{- \tttilde{g} \del^3_{x;1,2}\del_{y;1,2}}{(2\pi)^4} 
    \left[ \int d^4k \frac{e^{-\frac{k^2}{\Lam^2}}}{k^2+M_{\chi}^2}+  
       \int d^3k_xdk_y \frac{e^{ik_y(-2y+2y_0)-\frac{k^2}{\Lam^2}}}
                      {k_x^2+k_y^2+M_{\chi}^2}
    \right] ~.
\end{eqnarray}
Repeating the geometric series expansion in $k^2/M_{\chi}^2$,
\begin{eqnarray}
  \label{eq:48b}
\hspace{-35pt}
\xpv{\phi\phi}_{\chi} &=& \frac{-
  \tttilde{g}\del^3_{x;1,2}\del_{y;1,2}}{M_{\chi}^2(2\pi)^4} \times \non
&& \sum_{n=0}^{\infty}
    \left[ \int d^4k \left(\frac{-k^2}{M^2_{\chi}}\right)^n 
      e^{-\frac{k^2}{\Lam^2}} 
      +  
\int d^3k_xdk_y \left(\frac{-k_x^2-k^2_y}{M_{\chi}^2}\right)^n
    e^{ik_y(-2y+2y_0)-\frac{k^2}{\Lam^2}} 
    \right]. 
\end{eqnarray}
we see that we can extract 
the $k_y$ dependence in the second term as a
    derivative. The $x$ dependence {along} the boundary and the
    full bulk term give purely local
    corrections as expected from loop graphs. Though this non-local
    $y$-dependence is counterintuitive,
    the physical reason is easily identified. It is the interaction with
    the image charge. We find 
\begin{eqnarray}
  \label{eq:60b}
&& \hspace{-10pt}
  \xpv{\phi\phi}_{\chi} =
\non && =
{\rm bulk} + 
    \frac{- \tttilde{g} \del^3_{x;1,2}\del_{y;1,2}}{M_{\chi}^2(2\pi)^4} 
    \left[  \sum_{n=0}^{\infty}\sum_{p=0}^n \left(\matrix{n \cr
      p}\right) \left( \frac{\pa_y^2}{M^2_{\chi}} \right)^p 
      \int d^3k_xdk_y \left(\frac{-k_x^2}{M_{\chi}^2}\right)^{n-p}
      e^{ik_y(-2y+2y_0)-\frac{k^2}{\Lam^2}} 
    \right]  \non
&&=
  {\rm bulk} + 
    \frac{- \tttilde{g} \del^3_{x;1,2}\del_{y;1,2}\Lam^3}{M_{\chi}^2(2\pi)^4} 
    \left[  \sum_{n=0}^{\infty}\sum_{p=0}^n \alp_{n-p}\left(\matrix{n \cr
      p}\right) \left( \frac{\pa_y^2}{M^2_{\chi}} \right)^p 
      \int dk_y 
      e^{ik_y(-2y+2y_0)-\frac{k_y^2}{\Lam^2}} 
    \right]  \non
&&=
  {\rm bulk} + 
    \frac{- \tttilde{g} \del^3_{x;1,2}\del_{y;1,2}\Lam^3\pi^{1/2}}{M_{\chi}^2(2\pi)^4} 
    \left[  \sum_{n=0}^{\infty}\sum_{p=0}^n \alp_{n-p}\left(\matrix{n \cr
      p}\right) \left( \frac{\pa_y^2}{M^2_{\chi}}\right)^p \Lam 
e^{-\Lam^2(y-y_0)^2}    \right]~.
\end{eqnarray}
where $\alp_{n} = 2\pi^{3/2}(-2)^{n+1}(2n+1)!!$. 
In the distributional sense this is therefore equal to
\begin{eqnarray}
  \label{eq:61b}
  \xpv{\phi\phi}_{\chi} 
  &=&
  {\rm bulk} + 
    \frac{- \tttilde{g} \Lam^3}{M_{\chi}^2} 
    \left[ \sum_{p=0}^{\infty}\zeta_p\frac{\pa_y^{2p}}{M^{2p}_{\chi}}
  \del(y-y_0) 
    \right]~. 
\end{eqnarray}
where $\zeta_p$ can be read off from (\ref{eq:60b}).
The bulk one-loop $\chi$-diagrams therefore gives rise to the 
higher-derivative irrelevant corrections on the boundary
\begin{eqnarray}
  \label{eq:65}
  S_{eff} = \sum_p\oint d^3x  \frac{\tttilde{g}\beta_p\Lam^3}{M_{\chi}^2}\phi\left(\frac{\pa_n^{2p}}
{M_{\chi}^{2p}}\right)\phi~.
\end{eqnarray}
This result shows that the boundary irrelevant operators will
generically {not} appear in the combination $\oint
\pa_i\phi\pa_i\phi-\phi\pa_n^2\phi$. This is a direct consequence of
the fact that the boundary breaks Lorentz invariance. Examples which
generate the other two irrelevant operators are easily found. The
model just discussed will also generate $\oint \phi^4$ terms. A
non-linear sigma model will naturally have
$\oint \pa_n\phi\pa_n\phi$ corrections.

\subsubsection{Minkowski space boundary conditions and irrelevant operators}
\label{sec:mink-space-bound-1}

An important question therefore is how generic the 
occurrence of irrelevant
corrections is. In particular do fixed
points of boundary RG-flow, e.g. the Minkowski boundary conditions or
other `vacua', still receive irrelevant corrections? RG principles
tell us that we should expect them. Just because we are at a fixed
point of RG-flow, does not mean that irrelevant operators encoding 
a high-energy sector are forbidden.  In the context of boundary
RG-flow, the connection
between boundary conditions and `vacua', makes this statement somewhat
surprising. In Minkowski space in particular we do not expect that
integrating out a high-energy sector would change the vacuum state in
the low-energy effective theory even at the irrelevant
level.\footnote{We thank Jim Cline for emphasizing this point.} Both the
general RG principles and the intuition that in Minkowski space high
energy physics should not change the low-energy boundary conditions
are true, as we will now illustrate. The first point 
is evident from the two scalar 
theory 
at the beginning of this section with the interactions given in 
(\ref{eq:41}). Integrating out the $\chi$ field
exactly, clearly gives rise to the following irrelevant contributions to 
the low-energy effective theory for $\phi$.
\begin{eqnarray}
  \label{eq:19}
  S_{low-energy}^{int} &=& \hlf \int d^3xdy -
  \phi(g+\gam\del(y-y_0))(\Box_{bc_{\chi}} -M_{\chi}^2)\inv
  (g+\gam\del(y-y_0))\phi \non
  &=& {\rm bulk} + \sum_{n=0}^{\infty} \hlf \oint \frac{2\gam
  g}{M_{\chi}^2}\phi\left(\frac{\Box_{bc_{\chi}}}{M_{\chi}^2}\right)^n \phi
  + \frac{\gam^2}{M_{\chi}^2}\phi
  \left(\frac{\Box_{bc_{\chi}}}{M_{\chi}^2}\right)^n \del(0)\phi  ~.
\end{eqnarray}
Here $\Box_{bc_{\chi}}$ should be interpreted as acting on
a complete set of eigenfunctions with the boundary conditions
$\pa_n\chi = -\kap \chi$ that
belong to the massive field $\chi$. 
To address the formal divergence of the delta 
function at its origin, $\del(0)$, 
recall first that in a cut-off theory, as we are
considering, all distributions become smeared on the scale of the
cut-off. The $\del(0)$ in the second term is therefore
proportional to $M_{\chi}$ purely on dimensional grounds. Our cut-off
scheme eq. (\ref{eq:50}) indicates that $\del(x) = \lim_{\Lam \rar
  \infty}  \pi^{-1/2}\Lam e^{-\Lam^2x^2}$, $\del(0) =
M\pi^{-1/2}$. This regularization only postpones the problem,
however. In \cite{nous} we perform a computation, which indicates that
the $\del(0)$ term arising from discontinuous field redefinitions does
not explicitly appear in bulk correlation funcions. Its sole function is to
generalize all distributions so that they 
obey the correct boundary conditions $\pa_n f(y) = -\kap f(y)$.

Consistent with the principles of decoupling, we see that whatever
boundary conditions we choose for $\phi$ including fixed points of RG flow,
the boundary action will receive irrelevant corrections. How can this
possibly square with the idea that Minkowski space 
high energy physics should not correct
the vacuum choice, i.e. the Minkowski space 
boundary conditions of $\phi$? In this
simple model it is fairly easy to see that the boundary conditions of
$\phi$ change, because the massive field $\chi$ does {not} have
Minkowski space boundary conditions. 
When $\chi$ is integrated out, this reverberates in the low
energy effective boundary action for $\phi$. A naive way to see that
$\chi$ is not at a 
fixed point of boundary RG-flow, is to note that
 the full boundary condition for $\chi$ reads $\pa_n\chi =
 -\kap\chi-\gam\phi$. The explicit dependence on $\phi$ perturbs one
 away from a $\chi$-sector fixed point $\kap_{fixed}$. To consider a
 fixed point in the $\chi$-sector alone is inconsistent of course; the
 full $\chi$-$\phi$ dynamics needs to be taken into account. 
But an exact answer, possible because the theory is exactly solvable,
 shows that this naive guess is qualitatively correct. The exact
 answer is obtained by diagonalizing the theory to two fields $\Phi_1$
 and $\Phi_2$ with action
 \begin{eqnarray}
   \label{eq:20}
   S_{bulk} &=& \hlf \int d^3xdy\,\, \Phi_1\left( \Box - M_{\chi}^2 +
   \frac{g^2}{4M_{\Del}^2}\right)\Phi_1 + \Phi_2\left( \Box - m_{\phi}^2 +
   \frac{g^2}{4M_{\Del}^2}\right)\Phi_2 +\cO(g^3)~,
    \non
   S_{bound} &=& \hlf \oint d^3x \,\,\Phi_1\left(\frac{2g\gam}{M_{\Del}^2} +
   \frac{\gam^2 \del(0)}{M_{\Del}^2} \right)\Phi_1
   -\Phi_2\left(\frac{2g\gam}{M_{\Del}^2} +
   \frac{\gam^2 \del(0)}{M_{\Del}^2} \right)\Phi_2
   +\cO(\gam^3,g\gam^2,g^2\gam)~,
   \non
   M_{\Del}^2&=&M_{\chi}^2-m_{\phi}^2 ~.
\end{eqnarray}
If we tune $\gam$ and $g$ such that one of the two fields 
has
Minkowski boundary conditions $\kap_{\Phi_2} =
-i\ome(\vk,M_{\Phi_2})$, we see that the difference in masses
$M_{\Phi_1} \sim M_{\chi}$ and $M_{\Phi_2} \sim m_{\phi}$ prevents the
other from obeying Minkowski boundary conditions. 

At a very fundamental level these results are easily
understood. Recall that the Minkowski boundary conditions are the only
boundary conditions respecting Lorentz invariance; this is what
guarantees that the values of the boundary couplings
correspond to a fixed point. The explicit boundary
interaction 
$\oint -\gam\chi\phi \simeq -\hlf\int \del(y-y_0)\gam\chi\phi\,$ breaks Lorentz
invariance, however. In the diagonal system with $\Phi_1$, $\Phi_2$,
the Lorentz invariance is broken because one of the two fields does
not obey Minkowski boundary conditions. 

We have only shown that irrelevant operators will generically appear in a
situation where a field in the high energy sector is not in the
Minkowski vacuum. Lorentz symmetry should guarantee the converse: that
if all massive fields obey Minkowski boundary conditions, no boundary
RG-flow {or boundary
irrelevant operators} can appear. 
Importantly,
in the setting of interest
to us, FRW cosmology, Lorentz invariance is absent. 
It is therefore not clear that cosmological boundary conditions, 
to which we turn now, are similarly protected from RG-flow 
and irrelevant contributions from high energy physics.  
Strictly applying the RG principles, we should {\em not} expect them to be
protected.

\section{Boundary conditions in cosmological effective
  Lagrangians}
\label{sec:pref-bound-cond}
\setcounter{equation}{0}

We have seen that: 
\begin{itemize}
\item[(1)] a boundary action can encode the boundary
conditions one wishes to impose on the fields. 
\item[(2)] This 
holds in
full generality. The boundary need not correspond to a physical
obstruction or object. Completely transparent boundary conditions
exist that mimick the situation as if there is no boundary. 
{Introducing a boundary action to account for initial conditions
  therefore places no additional constraints on the theory.}
\item[(3)] Generically
the boundary conditions will be affected by 
RG flow, and suffer irrelevant corrections that are controlled by the
high energy physics.
\end{itemize}
We now use this knowledge to describe FRW cosmologies from a
Lagrangian point of view. The main issue in the Hamiltonian
description of FRW cosmologies is that of vacuum selection. In the
absence of a global time-like Killing vector or asymptotic flatness,
there is no unique vacuum state. There are two preferred candidates, the
Bunch-Davies and the set of adiabatic vacuum states, which we review below,
but some uncertainty remains. Both states, in fact, rely on an
asymptotic condition which ceases to be valid in the presence of a
finite Planck scale. We wish to emphasize, however, that whichever state is the true one, points (1) and
(2) above tell us that we can account for this state by the
introduction of a specific boundary condition at an arbitrary time
$t_0$.

Our lack of knowledge of the specifics of the very early universe and 
the high energy degrees of
freedom dominating at that time rather suggests to encode 
the initial state
uncertainty in a `past boundary' for any cosmological
theory. With the boundary comes the Lagrangian translation of the
vacuum choice ambiguity: what boundary conditions
to impose?  We will not give an answer to this long-standing
question. We will show, however, that whatever (local relevant) 
boundary conditions
one chooses, they are consistent in the sense that the backreaction is
under control. The 
counterterms {\em appropriate to the boundary conditions specified}
that are necessary to render the Minkowski
stress-tensor finite, do so in cosmological setting as well. This
confirms the intuition that the boundary conditions do not affect
UV-physics. And this continues to hold for any choice of cosmological initial
conditions. This may come as a surprise. The Hadamard condition
--- that at short distances 
the two-point correlation function is the appropriate
power of the geodesic distance 
$\sig(x_1,x_2)^{d-2}$ --- has long been thought to be a consistency
requirement for cosmological boundary conditions. Only these
correlation functions permit
`renormalization' by the standard Minkowski stress tensor. The lesson
from section \ref{sec:deco-theor-with}, however, is that other
short distance behavior does not necessarily 
signal an inconsistency, but instead implies
that the `boundary conditions' need to be renormalized as well. This
returns to the front the question which boundary conditions describe the
physics of the real world, but {\em none} that can be deduced from
local relevant boundary interactions are intrinsically
inconsistent. This is the power of the effective Lagrangian point of view.

Suppose for now that all choices for boundary conditions on the
initial surface of an FRW universe
are indeed consistent. 
Compared to Minkowski spacetime 
there is a new ingredient. 
The boundary condition needs to be covariantized. This is done by the
introduction of a unit vector $n^{\mu}$ normal to the boundary.
\begin{eqnarray}
  \label{eq:55}
  \pa_n\phi \equiv n^{\mu}\pa_{\mu\phi}
  =0~,~~~|g_{\mu\nu}n^{\mu}n^{\nu}| = 1.
\end{eqnarray}
In the conformal frame,
\begin{eqnarray}
\label{eq:77}
ds^2_{FRW}= a^2(\eta)(-d\eta^2+dx_{d-1}^2), 
\end{eqnarray}
the unit normal vector to the boundary 
scales as $a^{-1}$. Hence the boundary condition reads
\begin{eqnarray}
  \label{eq:49}
 \frac{1}{a} \pa_{\eta} \phi|_{\eta=\eta_0} = - \kap \phi|_{\eta=\eta_0}~.
\end{eqnarray}
The explicit dependence on the scale factor $a$ 
simply reflects that momenta redshift under cosmic expansion.\footnote{ 
Realizing that cosmological scaling induces
RG-flow we manifestly see the previous claim 
that Dirichlet conditions are trivial IR-fixed
points.} 
To construct the two-point correlation
function for a massive scalar $\phi$ that satisfies this boundary
condition, we need the equation of motion in an FRW background. For
simplicity we will assume that this background is pure de Sitter; the
results below generalize straightforwardly to power-law inflation and
are therefore truly generic. The equation of motion is  
\begin{eqnarray}
  \label{eq:51}
 && \frac{1}{\sqrt{-g}}\pa_{\mu}\sqrt{-g}g^{\mu\nu}\pa_{\nu}\phi(x,\eta)
  -m^2\phi(x,\eta) = 0~,  \non
&\Rightarrow& \left(\ove{a^2}\pa_{\eta}^2+
 (d-2)\frac{a'}{a^3}\pa_{\eta}+\frac{\vk^2}{a^2}+m^2\right)\phi(\vk,\eta) = 0~.
\end{eqnarray}
In the second step we Fourier transformed the spatial
directions. Substituting the constant de Sitter Hubble radius
$a^{-2}a' = H$, 
the explicit scale factor $a=-1/H\eta$ 
and making the
conventional redefinition $\eta=-y/\vk$, we have a Bessel equation for
$\tttilde{\phi} \equiv y^{-(d-1)/2}\phi$:
\begin{eqnarray}
  \label{eq:52}
\left(y^2\pa^2_y+y\pa_y+y^2+\frac{m^2}{H^2}-\frac{(d-1)^2}{4}
\right)\tttilde{\phi}(\vk,y)=0 ~.
\end{eqnarray}
The most general solution to the field equation is
therefore
\begin{eqnarray}
  \label{eq:53}
  \vphi_{b_{\kap}}(\vk,\eta)&=&
  \vphi_{dS,+}+b_{\kap}\vphi_{dS,-} \non
\vphi_{dS,+} &\equiv&(-\vk\eta)^{(d-1)/2} \sqrt{\frac{\pi}{4\vk}}\left(\frac{H}{\vk}\right)^{\frac{d-2}{2}} \bbbar{H}_{\nu}(-\vk\eta)~,~~~\nu = \sqrt{\frac{(d-1)^2}{4}-\frac{m^2}{H^2}}~,
\end{eqnarray}
with $H_{\nu}(y)$ the Hankel function satisfying eq.(\ref{eq:52}). 
The normalization and convention is such that in the limit $\vk \rar
\infty$ we recover the Minkowski space solutions. 
The boundary conditions (\ref{eq:49}) determine $b$, 
as in eq. (\ref{eq:15}). 

By construction the Green's function is given by\footnote{
\label{fn:3}
A `covariant' Green's function is given by 
  \begin{eqnarray}
\nonumber
    \label{eq:69}
    G_{\kap_f,\kap}(\vk_1,\eta_1;\vk_2,\eta_2) = (2\pi)^3\del^3(\vk_1+\vk_2)
    \sum_n^{trunc(\kap_f)} \mu(n)
    \frac{\phi_{b_{\kap},n}(\eta_1)\phi_{b_{\kap},n}(\eta_2)}{
    H^2n^2-m^2+H^2(d-1)^2/4}~. 
  \end{eqnarray}
where $\kap_f$ characterizes the future boundary condition at
$\eta=\infty$ and $\mu(n)$ is an easily determined measure.
From this expression it is clear that the 
delta function therefore also obeys the boundary condition. Indeed the
delta function is best viewed as a completeness relation for
eigenfunctions of the Laplacian $\Box \vphi_k = -k^2 \vphi$ obeying
$a_0\inv\pa_{\eta} \vphi_k|_{\eta_0} = -\kap\vphi_k|_{\eta_0}$, i.e.
\begin{eqnarray}
  \label{eq:81}
\nonumber
  \del_{\kap}(\eta_1-\eta_2) = \sum_n \mu(n)
  \phi_{b,n}(\eta_1)\bbbar{\vphi}_{b,n}(\eta_2) 
\end{eqnarray}
}
\begin{eqnarray}
  \label{eq:68}
  &&
\hspace{-0.3in}
G_{\kap_f,\kap}(\vk_1,\eta_1;\vk_2,\eta_2) =(2\pi)^3
  \del^3(\vk_1+\vk_2)\cN_{\kap_f,\kap}\left(\bbbar{\vphi}_{b_{\kap_f}}(\vk_1,\eta_1){\vphi}_{b_{\kap}}(\vk_2,\eta_2) 
  \theta(\eta_1-\eta_2) 
\right.
\non
&&
\hspace{2.6in}
\left.  
+
  {\vphi}_{b_{\kap}}(\vk_1,\eta_1)\bbbar{\vphi}_{b_{\kap_f}}(\vk_2,\eta_2)\theta(\eta_2-\eta_1)\right)  ~,
\end{eqnarray}
where $\kap_f$ characterizes the future boundary conditions at
$y=\infty$. The normalization $\cN_{\kap_f,\kap}$ is chosen such that 
$(\Box-m^2) G =
i\del^d/\sqrt{-g}$. 
This requires that
\begin{eqnarray}
  \label{eq:70}
\cN_{\kap_f,\kap} \vphi_{b_{\kap}}(\vk,\eta)\stackrel{\leftrightarrow}{\pa_{\eta}}
  \bbbar{\vphi}_{b_{\kap_f}}({\vk},\eta)  
  = -i a^{2-d}(\eta) = -i(-H\eta)^{d-2} ~.
\end{eqnarray}
We find that 
\begin{eqnarray}
  \label{eq:71}
  \cN_{\kap_f,\kap} =
  \frac{1}{(1-\bbbar{b}_{\kap_f}b_{\kap})}~.
\end{eqnarray}
From here on we will again restrict our attention to $d=4$ 
spacetime dimensions.

\subsection{Harmonic oscillator and shortest length boundary conditions}
\label{sec:bound-cond-adiab}

A special set of boundary conditions are the covariantization of the
completely transparent ``Min\-kow\-ski'' boundary conditions of
eq. (\ref{eq:59}).
We will call these ``harmonic oscillator'' boundary 
conditions.
Recall that these correspond to the 
boundary action $\oint\phi \sqrt{\pa_i^2-m^2}\phi$. Covariance 
requires that the scale factor should enter here as well. We thus find that 
the {\em cosmological} harmonic oscillator boundary condition is characterized by
\begin{eqnarray}
  \label{eq:8}
  \kap_{HO} = -i\sqrt{\frac{\vk^2}{a_0^2} +m^2}~.
\end{eqnarray}

For the specific momentum dependent 
choice of boundary location $\eta_0^{SL}(\vk) = -\Lam/H |\vk|$ or 
equivalently $a_0 = |\vk| / \Lambda$, these boundary
conditions correspond to a {\em constant} value for the physical
parameter $b$. They are therefore the 
boundary conditions proposed in
\cite{egks,Danielsson:2002kx}. 
Underlying this inspired choice is the thought
that in a cosmological theory there is an `earliest time', where a
physical momentum $p \equiv \vk/a(\eta)$ reaches the cut-off
scale (the shortest length). 
Whether there is truly an earliest time in cosmological
theories is an interesting question in its own right. It would be the
natural location for the boundary action, but as a consequence 
of the symmetry between boundary location $\eta_0$ 
and coupling $\kap$ exposed in section \ref{sec:freed-choice-bound}, 
it is not directly relevant to
us. Indeed it is easy to see that a momentum-independent 
coupling $\kap_{HO}$ at
$\eta^{SL}_0(\vk)=-\Lam/H|\vk|$ is equivalent to a boundary action on
a standard time-slice $\eta_0'$ with momentum-dependent 
coupling $\kap_{SL}$
\begin{eqnarray}
  \label{eq:85}
  \kap_{SL} &=& - \frac{\pa \phi_+(\eta_0') +b_{SL}
  \pa\phi_-(\eta_0') }{\phi_+(\eta_0')+b_{SL}\phi_-(\eta_0')}~,~~~~
  b_{SL} = -
  \frac{\kap_{HO}\phi_+(\eta_0^{SL})+\pa\phi_+(\eta_0^{SL})}
       {\kap_{HO}\phi_-(\eta_0^{SL})+\pa\phi_-(\eta_0^{SL})}~.
\end{eqnarray}

In the limit $\Lam \rar \infty$ we recover the harmonic oscillator vacuum at
$\eta=-\infty$. The coupling $\kap'$ encodes
these harmonic oscillator boundary conditions at $\eta_0=-\infty$ in terms of
conditions at $\eta_0'$ {\em plus} corrections that
vanish as $\Lam \rar \infty$. As we have seen in the previous section
and will discuss in detail in the next, these corrections therefore correspond
to the introduction of {\em specific irrelevant} boundary operators.

\subsection{The Bunch-Davies and adiabatic boundary conditions}

In universes without a global timelike Killing vector, there is no
clear concept of the vacuum as a lowest energy state. Particle number
is also not conserved and one cannot unambiguously define an `empty'
state either. Instead one must specify a particular in-state
characterizing the initial conditions. Two solutions to this
vacuum choice ambiguity have become preferred. One is the Bunch-Davies
vacuum, which is indirectly 
constructed by requiring that for high momenta $\vk/a
\gg H$ the Green's function reduces to the Minkowski one. The second
corresponds to the set of ($n$-th order) 
adiabatic vacua, which is constructed by the requirement that the
number operator on the 
vacuum changes as slowly as possible
\cite{BD,Chung:2003wn}.\footnote{Referring to our earlier comment, we
  see why the definitions of the BD and adiabatic state become
  ambiguous in the presence of a finite Planck scale. For the former
  the strict high $\vk$ limit does not exist ($\vk/a \leq
  M_{Planck}$). For the latter the adiabaticity parameter (roughly
  $aH/\vk$) is no longer arbitrarily small.}   
For de Sitter
space the infinite order vacuum and the Bunch-Davies one are the same;
we shall therefore only discuss the latter.

The boundary conditions corresponding to the
Bunch-Davies vacuum are readily found. 
In the basis (\ref{eq:53}) 
we have chosen, the Bunch-Davies-state corresponds
to choosing $b=0$, and hence
\begin{eqnarray}
  \label{eq:74}
  \kap_{BD}  &=& - \frac{\pa_n \vphi_{dS,+,0}}{\vphi_{dS,+,0}}~.
\end{eqnarray}
Note that the Bunch-Davies boundary conditions are the analogues of
the Minkowski boundary conditions in a {mathematical sense}
only. The flat space Minkowski boundary conditions in 
eq. (\ref{eq:86}) are easily recognized as 
$\kap_{Mink}^{flat-space}=
-\pa_n\vphi_{Mink,+,0}/{\vphi_{Mink,+,0}}$ with
$\vphi_{Mink,\pm}\simeq e^{\pm i\ome t}$. 
Using the Bessel function recursion relation 
\begin{eqnarray}
  \label{eq:66}
  \pa_y H_{\nu}(y) = \frac{\nu}{y}H_{\nu}(y) - H_{\nu+1}~,
\end{eqnarray}
and the chain rule $\pa_\eta = -\vk\pa_y$ (recall that
$\pa_n=a\inv\pa_{\eta}$) a straightforward calculation yields
\begin{eqnarray}
  \label{eq:67b} 
\kap_{BD} &=& -\frac{\vk}{a_0}\left(\frac{\bbbar{H}_{\nu+1}(-\vk{\eta}_0)}{\bbbar{H}_{\nu}(-\vk\eta_0)}
 +\frac{(d-1)+2\nu}{2\vk\eta_0}\right) \non
&=&
 -\frac{\vk}{a_0}\left(\frac{\bbbar{H}_{\nu+1}(-\vk\eta_0)}{\bbbar{H}_{\nu}(-\vk\eta_0)}\right)
  + H\frac{(d-1)+2\nu}{2}~.
\end{eqnarray}
Knowing the asymptotes of the Hankel functions
\begin{eqnarray}
  \label{eq:73}
z \rar 0 &:&  H_{\nu} (z) \sim
-i\frac{1}{\sin(\nu\pi)\Gam(1-\nu)}\left(\frac{2}{z}\right)^{\nu}  =
-i\frac{\Gam(\nu)}{\pi}\left(\frac{2}{z}\right)^{\nu}~,\\ 
z \rar \infty &:& H_{\nu} (z) \sim \sqrt{\frac{2}{\pi z}} e^{i(z-\hlf
  \nu\pi -\frac{1}{4}\pi)}~,
\end{eqnarray}
we see that for $\eta_0 \rar -\infty$ the Bunch-Davies boundary
condition reduces to harmonic oscillator
boundary conditions 
\begin{eqnarray}
\kap_{BD} &\simeq& -\frac{|\vk|}{a_0}\left(  e^{\frac{i\pi}{2}} \right)
  +H\frac{(d-1)+2\nu}{2} \non 
 &\simeq& - i \frac{|\vk|}{a_0}
\end{eqnarray}
of a massless field. (One cannot say that the boundary
conditions tend to Dirichlet, the diverging $a_0$ is compensated by
the normal vector, see eq. (\ref{eq:49}).) The mass correction is
subleading in this limit.
We should keep in mind though that this is a formal expression. At
$\eta_0=-\infty$ the induced boundary volume vanishes, and boundary
conditions cannot easily be accounted for in terms of a boundary
action.

\subsection{Transparent, thermal, adiabatic boundary
  conditions; fixed points of boundary RG
flow?}

The most natural choice for the boundary conditions are arguably the
ones which are {transparent}. If there is no real
interface at the boundary location $y_0$, no physical effects of its
location should be 
noticeable.
To define transparency we need a notion of incoming and outgoing
waves. A clean definition of such waves only exists in asymptotically
flat spaces. Suppose one establishes these and let us call the incoming
wave (from the past) $\vphi_-$ and the outgoing $\vphi_+$. 
The transparent boundary
conditions are then those with $b_{\kap}=0$. Of course de Sitter space is not
asymptotically flat, but based on the asymptotic behavior of
the Bessel functions, 
one can argue that the basis
functions $\vphi_{dS,-}$ and $\vphi_{dS,+}$ defined in (\ref{eq:53})
correspond to in- and out-going waves respectively. 
In that sense the
Bunch-Davies boundary conditions are the transparent ones.

A definition which is more intrinsic to de Sitter is that the
Bunch-Davies boundary conditions are the thermal
boundary conditions. This emphasizes the existence of a cosmological
horizon, and is probably tied to the notion of transparency.  From the
Lagrangian point of view the true vacuum should be a (UV) fixed point
of boundary RG-flow. 
In the presence of a global timelike Killing vector with a conserved
quantum number $\pa_{t}\phi
= iE\phi$ such a fixed point is easily constructed following the
Minkowski space example in section \ref{sec:mink-space-bound}. In
cosmological spacetimes it is not clear what the fixed points of
boundary RG-flow are or whether there are any. The absence of a unique
vacuum suggests that there may be none. If we recall that cosmological
expansion induces RG-flow, the definition of the adiabatic vacuum, i.e.
that the number operator on the vacuum change as slowly as possible,
becomes very interesting. 
It would be worthwhile to investigate 
these connections between the transparent (i.e. 
Bunch-Davies), the thermal, and the adiabatic vacuum in FRW 
backgrounds and fixed points of boundary RG-flow further. 

\subsection{Backreaction and renormalizability for arbitrary boundary
  conditions} 
\label{sec:backr-renorm-arbitr}

We shall now make a crucial point. Any cosmological 
boundary condition $\kap$,
provided it is a dimension-one analytic function of the spatial
momenta, is consistent in the sense that backreaction is under
control. The divergences appearing in the stress tensor must, of
course, be regulated by the flat space counterterms of the {\em same}
theory. This includes the boundary counterterms for $\oint \kap
\phi^2$ and $\oint \mu\phi\pa_n\phi$. Our review in section
\ref{sec:deco-theor-with} has made this clear. In a rather coarse
fashion we can also see this directly from the FRW Green's function in
the limit of high (spatial) momentum --- in as far as this limit exists in a
cut-off theory. Using the
asymptotic values of
the Hankel functions, the basis functions
$\phi_{\pm,dS}(\vk,\eta)$ tend to massless Minkowski ones (the mass is
negligible in the high momentum limit)
\begin{eqnarray}
  \label{eq:80}
  \vk \rar \infty &:& \phi_{\pm,dS}(\vk,\eta) \simeq 
  \ove{\sqrt{2\vk}}\frac{e^{\pm i \vk \eta}}{a}  =
  \frac{\phi_{\pm,Mink}(\vk,\eta)}{a} ~.
\end{eqnarray}
The 
coefficient $b$ encoding the effective boundary
conditions for high-momentum modes 
therefore does {not} vanish, but reads 
\begin{eqnarray}
  \label{eq:12}
  b &=& - \frac{ \kap \phi_{+,Mink,0}+a_0\inv\pa_{\eta}\phi_{+,Mink,0} -
  H\phi_{+,Mink,0}}
{ \kap \phi_{-,Mink,0}+a_0\inv\pa_{\eta}\phi_{-,Mink,0} -
  H\phi_{-,Mink,0}} \non
 &=& -\frac{a_0\kap +i|\vk| +a_0H}{a_0\kap-i|\vk|+a_0H}e^{2i|\vk| \eta_0}~.
\end{eqnarray}
The last terms in 
the numerator and 
the denominator are 
negligible in this limit
$|\vk| \gg aH$. They are 
  remnants of the fact that the background breaks Lorentz invariance. 
The coefficient $b$ thus does {not}
vanish in the high momentum limit. Because a non-zero $b$ means that
there will be divergences in the theory {\em aside} from the
`Minkowski'-space divergences, it appears that any choice of boundary
conditions with $b\neq 0$ is in trouble. In section
\ref{sec:deco-theor-with} we reviewed, however, that this is not
so. The additional divergences are {localized} on the boundary
surface where the boundary conditions are imposed, and can be
reabsorbed in a redefinition of the {boundary} couplings. Any
choice for $b$ (descending from a boundary coupling $\kap$ that is
dimension one and analytic in the spatial momenta) is consistent.

One is tempted to conclude 
that for any boundary condition imposed at $\eta_0 =
-\infty$, the high spatial momentum limit of $b$ vanishes. 
This is true in the sense that if we keep $\kap$ fixed 
our flat space intuition, that boundary effects
vanish when the boundary is moved off to infinity, continues to hold. 
However, this goes against the principles behind the framework we
advocate here. In the sense of the symmetry between boundary location and
boundary coupling $\kap$, as explained in section
\ref{sec:freed-choice-bound}, it is only the specific 
combination $b_{\kap}$ which matters. At what location $\eta_0$ one imposes
the boundary conditions $\kap$ is immaterial to the physics.

The conclusion is that the answer to the question ``what
boundary conditions should we impose on quantum fields in FRW backgrounds''
requires physics input rather than internal consistency. 
The
Bunch-Davies vacuum certainly seems the closest analogue of Minkowski
boundary conditions,  even though it is not the naive covariantization
of them. 
The similarity suggests that the Bunch-Davies boundary conditions
may correspond to a fixed point
of boundary RG-flow. At the same time Lorentz symmetry is still
broken. If they are renormalized, it would 
suggest that they are not special in any
sense.

Let us emphasize again that we have shown consistency, i.e. a
manifestly finite backreaction. The observed energy density of our
current universe will or will not agree with the predictions for the
backreaction based on using different boundary conditions. 
This, however, is precisely the physics input that is needed. Only a
measurement can decide the correct boundary conditions to be used in
any situation.

\section{Transplanckian effects in Inflation}
\label{sec:transpl-effects-infl}
\setcounter{equation}{0}

Inflationary cosmologies are the leading candidates to solve the
horizon and flatness problems of the Standard Model of Cosmology. 
Consistency with the observed spectrum of temperature fluctuations in
the Cosmic Microwave Background (CMB) provides an estimate of the Hubble parameter $H$
during inflation. Depending on the model, 
$H$ can be as high as
$10^{14}$ GeV. With the string scale \mbox{$M_{string}=10^{16}$ GeV} as
the scale of new physics, this means that 
the suppression factor $H/M$ of 
irrelevant operators could optimistically be 
at the one-percent level.
This opens a window of opportunity to {\em
  experimentally witness}
effects of Planck scale physics \cite{Brandenberger}. 
Besides its theoretical appeal, inflation
is also the leading candidate for early universe cosmology 
on experimental grounds. The most
precise cosmological measurements to date, the temperature
fluctuations in the CMB, advocate inflation.
The CMB measurements are therefore also the most promising arena where
remnants of transplanckian physics could show up.
In inflationary cosmologies the CMB
temperature fluctuations originate in quantum fluctuations during
the inflationary era. The issue of vacuum selection in cosmological
settings thus has immediate consequences for CMB predictions. At the
classical level the Bunch-Davies choice is, for reasons reviewed in
the previous section, the preferred one; 
it is the closest analogue to
the Minkowski boundary conditions.
Previous investigations into effects of Planck scale physics 
suggest that the CMB
fluctuation spectrum is affected at leading order in $H/M_{Planck}$ and that
this effect is precisely due to the choice of vacuum
\cite{egks,Danielsson:2002kx}.  
Due to our ignorance of the details of Planck scale physics
(i.e. our lack of understanding of string theory in time-dependent
settings), decoupling in effective field theory is arguably the
framework in which transplanckian 
corrections must ultimately be understood \cite{stanford}. By the
addition of an arbitrary 
boundary action encoding the boundary conditions, 
we have put the issue of vacuum
selection on a consistent footing with the ideas of effective field
theory. In this comprehensive formulation, we can deduce systematically
what the effect of Planck scale physics is on 
boundary conditions (vacuum selection) and whether its effect on CMB
predictions is indeed leading
compared to bulk corrections.\footnote{The object of our study is an
external scalar field in a fixed FRW background. Strictly speaking
only  
the gravitational tensor fluctuations are effectively described by
such a model.  
However, our arguments should apply to the scalar-metric fluctuations
as well, 
since these only differ by an amplification factor of the inverse
slow-roll parameter.}

The Planck scale physics is encoded in
irrelevant operators. The leading bulk irrelevant operator
$\frac{1}{M^2}\int \phi \Box^2\phi$ consistent with the symmetries 
is dimension six. In section \ref{sec:wilsonian-rg-flow} we
constructed and derived 
the four leading irrelevant boundary operators in flat space 
\begin{eqnarray}
  \label{eq:42b}
  \ove{M}\oint_{y=y_0}\hspace{-.2in}d^3x\,\, \phi^4~,~~~
  \ove{M}\oint_{y=y_0}\hspace{-.2in}d^3x\,\, \pa^{i}\phi\pa_{i}\phi~,~~~ 
  \ove{M}\oint_{y=y_0}\hspace{-.2in}d^3x\,\, \pa_n\phi\pa_n\phi~,~~~ 
  \ove{M}\oint_{y=y_0}\hspace{-.2in}d^3x\,\, \phi\pa_n\pa_n\phi~.
\end{eqnarray}
compatible with unbroken $ISO(3)$ symmetry.
In a cosmological setting this is the 
requirement of homogeneity and isotropy.
These operators are all dimension four and as the explicit scaling
shows, they are expected to be dominant over the leading bulk
irrelevant operator. In curved space these operators
are covariantized. 
For a scalar field $\phi$ covariantization has only a significant effect on the
last operator in (\ref{eq:42b}). 
A new coupling is needed 
which provides the connection
for the covariant normal derivative
\begin{eqnarray}
  \label{eq:1}
   \ove{M} \oint \sqrt{h} n^{\mu}n^{\nu} \left(\phi \pa_{\mu}\pa_{\nu}
   \phi - \phi\Gam^{\rho}_{~\mu\nu}\pa_{\rho}(g) \phi \right)  = \ove{M}
   \oint \sqrt{h} n^{\mu}n^{\nu}D_{\mu}\pa_{\nu} \phi~.
\end{eqnarray}
Here $h_{ij} = g_{\mu\nu} \pa_i x^{\mu}\pa_j x^{\nu}$ is the induced
metric on the boundary, and $n^{\mu}$ its unit normal
vector. 
In FRW
cosmology with the metric in the conformal gauge,
\begin{eqnarray}
  \label{eq:21}
  ds^2_{FRW} = a^2(\eta)(-d\eta^2+dx_3^2)~,
\end{eqnarray}
and an
initial timeslice $\eta=\eta_0$ as boundary, the induced metric, connection
coefficients, and normal vector are 
\begin{eqnarray}
  \label{eq:2}
  h_{ij} &=& a^2_0(\delta_{ij})~,\non
  n^{\mu} &=& a_0\inv \delta^{\mu}_{\eta}~,\non 
\Gam^{\eta}_{~ij} &=& a_0 H_0\delta_{ij}~,~~ \Gam^i_{~\eta j} =
   a_0H_0\del^i_j~,~~ \Gam^{\eta}_{~\eta\eta} = a_0H_0~.
\end{eqnarray}
Here $a_0 \equiv a(\eta_0)$ and $H_0=H(\eta_0)$ is the Hubble radius
$H=a^{-2}\pa_{\eta}a$ at $\eta=\eta_0$. Substituting these values 
we obtain the FRW version of the irrelevant operator
\begin{eqnarray}
  \label{eq:84}
  \ove{M}\oint a_0^3 \phi \left(\pa_n - H\right)\pa_n\phi~.
\end{eqnarray}

We shall compute the effect of the leading irrelevant operators on the
two-point correlator of $\phi$. In inflationary cosmologies, the
latter determines the power 
spectrum of CMB density perturbations. We will assume we can treat the four-point
bulk $\lam\phi^4$ and (irrelevant) boundary interaction $\oint \phi^4$
perturbatively and will ignore them to first order. 
Combining the remaining irrelevant boundary operators in a correction
to the FRW boundary 
action, one obtains
\begin{eqnarray}
  \label{eq:23}
  S_{bound}^{irr.op.} = \oint_{\eta=\eta_0} d^3x a_0 \left[
  -\frac{\bpe}{2M}\pa^i\phi\pa_i\phi - \frac{\bpa}{2M}
  \pa_{\eta}\phi\pa_{\eta}\phi -\frac{\bco}{2M}{\phi}D_{\eta}\pa_{\eta}\phi\right]~.
\end{eqnarray}
The precise value of a coupling constants $\beta_i$ is determined by
{\em two} parts. (1) It is determined by
the details of the transplanckian physics; e.g. if transplanckian
physics is a free sector, decoupling is exact and $\beta =0$ (for
dynamical gravity the sectors are never decoupled of 
course),
 but (2) the couplings
$\beta_i$ are also {covariant} under the symmetry between boundary
location and coupling. If we would have computed the irrelevant
corrections to a boundary condition at a different location $y_0'$, we
would have found different values $\beta_i$ which upheld that all physical
quantities only depend on the choice of boundary location through a specific
combination $b_{\kap,\bet_i}$.
 
Two of the operators in eq. (\ref{eq:23}) contain normal derivatives. 
As discussed in section \ref{sec:deco-theor-with}, such operators can be
removed by a discontinuous field redefinition and a change of the
remaining boundary couplings. Doing so \cite{nous} we find that to lowest order in $\bet_i/M$, eq. (\ref{eq:23}) is  
equivalent to a boundary interaction (if the boundary coupling $\mu$=0)
\begin{eqnarray}
  \label{eq:4}
  S^{irr,leading} &=& \oint a_0^3 d^3x \, -\frac{\phi^2}{2}\left[\frac{\vk_1^2(\bpa-\bco)}{a_0^2M}+ \frac{\kap^2\bpe}{M}
  -\frac{\bco m^2}{M}-\kap\frac{3\bco H}{M}
\right]~,
\end{eqnarray}
where $m^2$ is the mass of the scalar field.
Fourier transforming along the boundary, 
the leading irrelevant correction thus amounts to a
change in the boundary condition $\kap$ by\footnote{Because the
  coupling $\kap$ is subject to renormalization, its  
value is fixed by a renormalization condition and an experimental
measurement. An important question therefore is, 
whether the effects of irrelevant operators
are experimentally measurable. The standard story, that (1) measured couplings
always include all relevant and irrelevant corrections, and that (2)
the contribution of each coupling $\beta_i$ is an independent contribution to  
the precise running of coupling $\kap_{eff}(\bet_i)$ 
under RG-flow, should apply. A very
precise measurement of the scaling behaviour of $\kap$ should reveal
the contributions of high energy physics encoded in the irrelevant
operators. This is explained in detail in the next subsection \ref{sec:an-earliest-time}.} 
\begin{eqnarray}
  \label{eq:5}
  \kap_{eff} =\kap_0 +\frac{\vk_1^2(\bpa-\bco)}{a_0^2M}+ \frac{\kap_0^2\bpe}{M}
  -\frac{\bco m^2}{M}-\kap_0\frac{3\bco H}{M}~.
\end{eqnarray}
We clearly see that the leading correction to the low-energy effective
action occurs at order $|\vk|/a_0M$ and $H/M$. For CMB physics the
momentum scale of interest is $|\vk|/a_{hor.crossing}\sim H$, and both are
of the same order. The conclusion that the $|\vk|$
dependent operators are suppressed by a factor $a_0/a_{hor.crossing}$ 
is
incorrect, when we recall that the location of the boundary is
arbitrary. 

For a given FRW universe the Green's function, including the $H/M$
correction to  
the boundary condition, can now simply be read off from
eqs. (\ref{eq:53})-(\ref{eq:68}).  
We can thus straightforwardly compute the leading transplanckian effect
on the power spectrum of inflationary perturbations. The latter is
related to the equal time Green's function with $\kap_f=\bar{\kap}$
\begin{eqnarray}
  \label{eq:6}
  P(\vk)_{\kap} &=& \lim_{\eta \rar 0} \frac{\vk^3}{2\pi^2}
  G_{\kap_f=\bar{\kap},\kap}(\vk,\eta;-\vk,\eta) \non
  &=& \lim_{\eta \rar 0} \frac{\vk^3}{2\pi^2} \frac{|\vphi_{b_{\kap}}(\vk,\eta)|^2}{(1-|b_\kap|^2)}~,
\end{eqnarray}
where $\vphi_{b_{\kap}}(\vk,\eta)$ is a solution to the (free) equation of
motion, normalized according to the inner product (\ref{eq:70}), and
with boundary condition $\pa_n\vphi| =-\kap\vphi|$. 
{\em Note that the
basis functions $\vphi_{b_{\kap}}$ only depend 
on the location of the 
  boundary through the physical combination $b_{\kap}$. This
  `independence' of the location of the boundary 
guarantees that the power-spectrum --- a physical quantity
  --- is so as well.} 
For an infinitesimal change in the boundary condition $\kap$, we can
treat the vertex $\oint -\hlf \del\kap \phi^2$ perturbatively, and the
change in the power spectrum simply amounts to computing the following
Feynman diagram.
\begin{eqnarray}
  \label{eq:47b}
    \SetScale{0.8}
     \begin{picture}(300,50)(220,5)
%  Second Graph
   \Line(400,45)(450,25)
   \Line(450,25)(500,45)
   \SetColor{Gray}
   \Line(450,27)(460,23)
   \Line(460,27)(470,23)
   \Line(470,27)(480,23)
   \Line(480,27)(490,23)
   \Line(490,27)(500,23)
   \Line(500,27)(510,23)
   \Line(510,27)(520,23)
   \Line(370,27)(380,23)
   \Line(380,27)(390,23)
   \Line(390,27)(400,23)
   \Line(400,27)(410,23)
   \Line(410,27)(420,23)
   \Line(420,27)(430,23)
   \Line(430,27)(440,23)
   \Line(440,27)(450,23)
%   \GBox(170,23)(330,27){.9}
   \SetColor{Black}
   \DashLine(450,25)(450,0){5}
  \end{picture}
\end{eqnarray}
This immediately illustrates that if $\del\kap$ is of order $H/M$, the
change in the power spectrum will be of order $H/M$. For completeness,
we compute the power spectrum by de Sitter Feynman diagrams in \cite{nous}. With the effective change in $\kap$
corresponding to the contributions of the irrelevant operators
$\bet_i$ known, we can also simply expand the exact solution for 
the power spectrum for any $\kap$. 
Choosing the
Hankel functions as basis as in eq. (\ref{eq:53}),
the solutions $\vphi_{b_{\kap}}$ are given by
\begin{eqnarray}
  \label{eq:9}
  \vphi_{b_{\kap}} &=&
  \vphi_{+}+b_{\kap}{\vphi}_{-}~,~~b_{\kap}=-\frac{\kap
  \vphi_{+,0}+ \pa_n\vphi_{+,0}}{\kap\phi_{-,0}+\pa_n{\phi}_{-,0}}~.
\end{eqnarray}
For an infinitesimal shift $\del\kap$ the 
power spectrum is thus
\begin{eqnarray}
  \label{eq:10}
 P(\vk)_{\kap+\del\kap} &=& P(\vk)_{\kap} 
  + \lim_{\eta \rar 0} \frac{\vk^3}{2\pi^2}\left[
    \frac{\del b}{(1-|b|^2)^2}
    \bbbar{\vphi}_{b_{\kap}}^2 
 + {\rm c.c.}
  \right] + \cO(\del b^2)~.
\end{eqnarray}
Substituting
the de Sitter values computed in the previous section, and using that 
asymptotically (see (\ref{eq:73}))
\begin{eqnarray}
\label{eq:11}
\lim_{\eta \rar 0} \bbbar{\vphi}_{b_{\kap},dS} =\frac{(1-\bbbar{b})}{(b-1)}\lim_{\eta \rar 0}\vphi_{b_{\kap},dS} ~,
\end{eqnarray}
we find that 
\begin{eqnarray}
  \label{eq:7}
 P(\vk)_{\kap+\del\kap} &=& P_{\kap} \left(1 +
 \frac{1}{(1-|b|^2)^2}\left[\del b\frac{(1-\bbbar{b})}{(b-1)}
 +{\rm c.c.} \right]
\right)~.
\end{eqnarray}
Recall from eq. (\ref{eq:14}) that 
\begin{eqnarray}
  \label{eq:13}
  \del b = -\frac{\del \kap
  \vphi_{+,0}}{\kap\vphi_{-,0}+\pa_n\vphi_{-,0}}+\frac{\del\kap\vphi_{-,0}(\kap\vphi_{+,0}+\pa\vphi_{+,0})}{(\kap\vphi_{-,0}+\pa_n\vphi_{-,0})^2} ~.
\end{eqnarray}
We see explicitly that the change in the power spectrum is also linear in 
$H/M$. 

For the preferred Bunch-Davies vacuum choice, where $b=0$, the
corrections thus become 
\begin{eqnarray}
  \label{eq:16}
    P_{BD+\del \kap}(\vk) &=& P_{BD}\left(1 + \left[\del \kap
    \frac{\vphi^2_{+,0}}{-\phi_{-,0}\pa_n\phi_{+,0}
    +\phi_{+,0}\pa_n\phi_{-,0}} +{\rm c.c.}\right] 
    \right)~.
\end{eqnarray}
It appears we have introduced a dependence on the boundary location,
but we should not forget that $\del\kap$ intrinsically depends on
$y_0$ as well. The combination above is guaranteed to be 
{independent} of the boundary location.
We recognize in the denominator the normalization condition
(\ref{eq:70}) (with $\pa_n=a\inv\pa_\eta$). The expression therefore
simplifies to
\begin{eqnarray}
\label{eq:17}
P_{BD+\del \kap}&=& P_{BD} \left(1+\left[\del\kap\frac{\phi^2_{+,0}}{-ia_0^{-3}} +{\rm c.c.}\right]+ \cO(\del\kap^2)\right)~.
\end{eqnarray}
Restricting our attention to de Sitter space, 
we insert the explicit expressions for the basis functions $\phi_+$
from eq. (\ref{eq:53}), and obtain, using that $a_0 = \vk/Hy_0$,
\begin{eqnarray}
  \label{eq:18}
  P_{BD+\del\kap}^{dS} &=& P_{BD}^{dS}\left( 1 -
\left(\frac{\pi}{4H}\right)
  \left[\frac{\del\kap\bbbar{H}_{\nu}^2(y_0)}{i} +{\rm c.c}\right]
  \right)~.
\end{eqnarray}
Substituting the irrelevant operator induced $\del\kap$ from
eq.~(\ref{eq:5}), we compute the following corrections to the power spectrum
\begin{eqnarray}
  \label{eq:75}
 && \hspace{-10pt}
{P_{\scriptscriptstyle BD+\del\kap}^{\scriptscriptstyle dS}} \!=
%  \non
% && 
P_{\scriptscriptstyle BD}^{\scriptscriptstyle dS}\left( 1 -\frac{\pi }{4H}
  \left[\frac{\bbbar{H}_{\nu}^2(y_0)}{i}
\left[\frac{\vk_1^2(\bpa-\bco)}{a_0^2M}+ \frac{\kap_{BD}^2\bpe}{M}
  -\frac{\bco m^2}{M}-\kap_{BD}\frac{3\bco H}{M}
\right] +{\rm c.c.}\right]
  \right),
\non
\end{eqnarray}
with (eq. (\ref{eq:67b}))
\begin{eqnarray}
\kap_{BD} &=& \frac{d-1+2\nu}{2} H -\frac{\vk}{a_0}
\frac{\bbbar{H}_{\nu+1}(y_0)}{\bbbar{H}_{\nu}(y_0)}~.
\end{eqnarray}
This is our final result. Let us stress again, that the apparent
dependence on the boundary location is only that. The boundary
couplings $\beta_i$ by
construction compensate 
the $y_0$ dependence
and the whole expression is independent of $y_0$.

\subsection{An earliest time in cosmological effective actions.\\ The
  inflationary power spectrum}
\label{sec:an-earliest-time}

We have repeatedly stressed that the location where one sets the
boundary conditions is immaterial. To compare the theoretical
predictions with experiment one must of course choose a specific
moment. Naively in cosmological spacetimes with a past singularity,
there is an `earliest time' which would be the logical candidate. We
will show here that the boundary effective action supplies a
`mathematical manifestion' of the concept of an `earliest time'. It
will be very clear, however, that this `earliest time' is an observer
dependent choice. The existence of the shift-symmetry is therefore
essential for consistency.\footnote{The results in this section were
  obtained together with B.R. Greene \cite{br}.}
 
Perturbative effective actions are intrinsically limited in their
range of validity to scales below the physical cut-off $M$. In an FRW
cosmology, this is manifest in the momentum expansion of the bulk
low energy effective action. The metric contributes a scale factor, so
that the small parameter is precisely the ratio of the physical
momentum to the cutoff: $\vk/a(t)M = p_{phys}(t)/M$. What is novel 
for cosmological effective actions is that the
boundary effective action parametrizing the initial conditions is an
expansion in the {\em blueshifted} momentum: It is in terms of the
physical momentum at the time where the initial conditions are set.
$\vk/a_0M = p_{phys}(t_0)/M$. The momentum expansion therefore has not
one but two small parameters and breaks down when either
\begin{eqnarray}
  \label{eq:5c}
  \frac{\vk}{a_0M} =1 ~~{\rm or}~~ \frac{\vk}{a(t)M} = 1~.
\end{eqnarray}
Physically this bounds mean the following.
If the physical processes we are are interested occur at co-moving 
momentum
scales ${\mu}_{co}$, then we immediately see that an FRW effective
action is only valid up to the `scale'
\begin{eqnarray}
  \label{eq:2c}
 \mu_{phys}(t) \equiv \frac{\vec{\mu}}{a(t)} = M~,
\end{eqnarray}
as is conventional, but it is also only valid up to a `time'
\begin{eqnarray}
  \label{eq:1c}
  a_0 = \mu_{co}/M~.
\end{eqnarray}
We see here the confirmation of our intuition that we can
only trust low energy effective cosmological 
theories up to the `Planck time'. So
far this has always been lacking.

\begin{figure}[htbp]
  \centerline{
\includegraphics[width=3.1in]{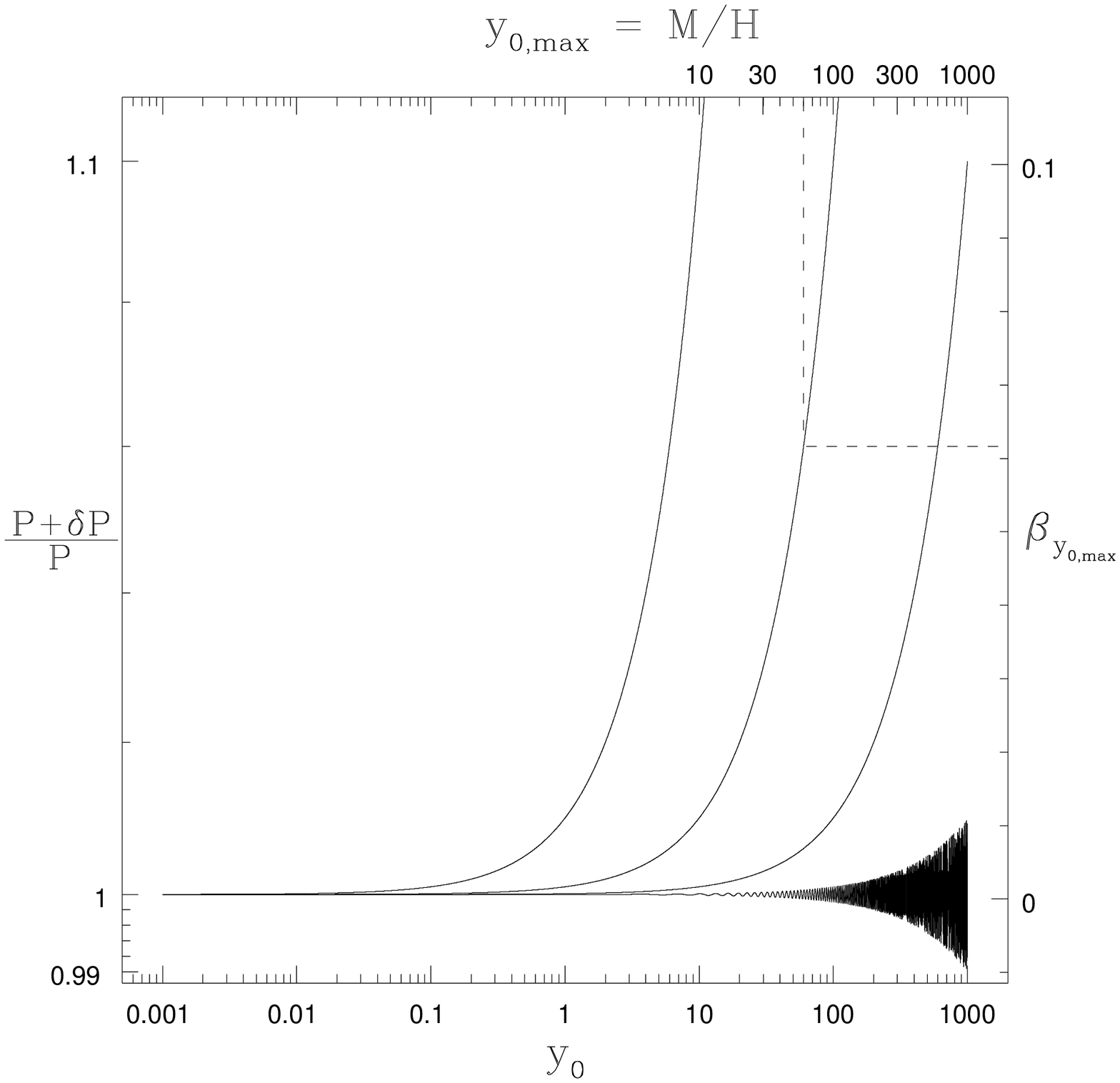}
\includegraphics[width=3.1in]{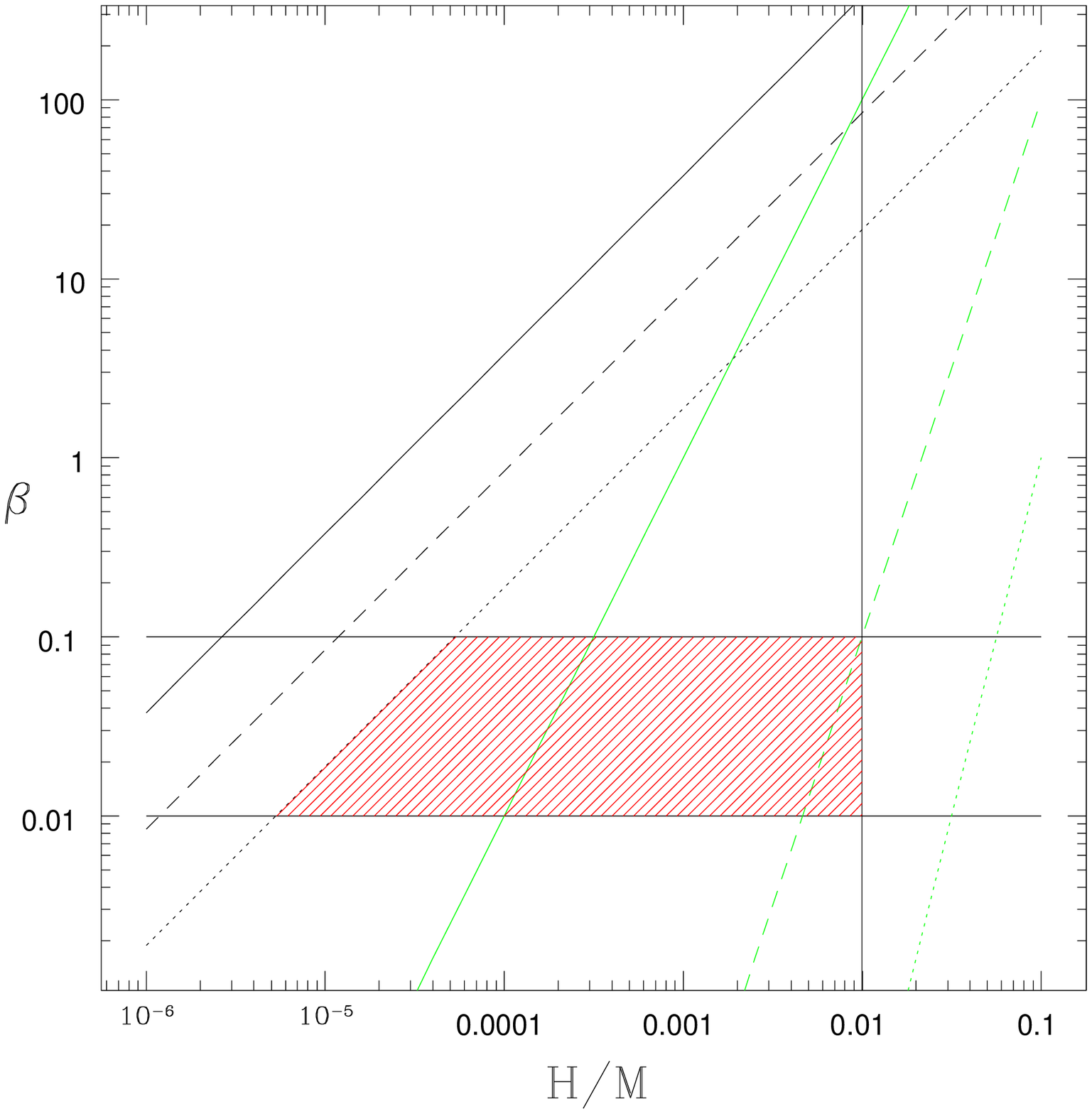}
}  
  \caption{
The left panel shows the change in the (amplitude of the) 
power spectrum  due to the
presence of the leading order irrelevant operator $\frac{\beta}{M}
(\pa_i\phi)^2$ as a function of
the physical momentum in units of the size of the horizon at the
'earliest time'. (Only for one specific choice is the full oscillatory
Bessel function behaviour plotted.) 
This graph should be read as follows. Given the scale
of new physics $M$ and the Hubble constant $H$ during inflation (or
more precisely at
the time when the highest mode $k_{max}$ of interest exits the horizon) the
earliest time up to which we can trust the effective action is when
$y_{0,max}\equiv k_{max}/a_{0,min}H =M/H$
(see subsection). Anything to the right of
$y_{0,max}$ should be discarded as untrustworty. Precisely at
$y_{0,max}$ the change in the power spectrum is linearly dependent on
the value of $\beta$. The values of $M/H$ and $\beta$ corresponding to the
various curves can thus be read of from the intersection of the
plumblines to the upper and right axis.
The right panel shows an exclusion plot for $\beta$ as a function of
$H/M$. The $45^o$ lines (black) correspond to the backreaction bounds (\ref{eq:260a})-~(\ref{eq:12c}) (continuous for zeroth order in slow
roll, dashed for first order in slow roll, dotted for second order in
slow roll). The $60^o$ lines (green) 
correspond to the order of magnitude
estimate made in \cite{Porrati:2004gz}. 
The upper horizontal line is an order of magnitude
estimation of the current error to which we have a nearly 
scale invariant spectrum \cite{Bennett:2003bz}. The lower horizontal line is an
order of magnitude estimate of the cosmic variance limitations of
resolution. Finally the vertical line denotes a maximal value of
$H/M$ consistent with observation. $H/M_{Planck}$ is extracted from the observed
amplitude of the power spectrum and we have set 
$M \equiv 10^{16}$ GeV. This leaves the
shaded region as the {\em window of opportunity} to observe
transplanckian physics in the CMB.
}
\label{fig:2}
\end{figure}

As stated, this 'earliest time' is then of course the logical place to locate to
boundary action to set the initial conditions. Doing so, we can
refine our analysis for which values of $\beta_i$ and $H/M$ changes in
the power spectrum are of the right order of magnitude to be
potentially observable. Note that for high $\vk$ all irrelevant
boundary operators reduce to a single one 
\begin{eqnarray}
  \label{eq:3}
  S^{irr,~leading,~high~\vk} = \oint a_0^3d^3x \, - \frac{\beta}{2M} \frac{\vk^2}{a_0^2}\phi^2
\end{eqnarray}
where $\beta=\beta_{\parallel}-\beta_{\perp}-\beta_c$. We will focus
on this single one for simplicity. This operator induces a correction
to the power spectrum of a massless field ($\nu=3/2$)
\begin{eqnarray}
  \label{eq:22}
  \frac{P_{BD}^{dS}+\delta P}{P^{dS}_{BD}}(y_0) = 1+
  \frac{\pi}{4}\frac{\beta H}{M} \left[ i y_0^2
  \bar{H}_{3/2}^2(y_0)+{\rm c.c.}\right]
\end{eqnarray}
The maximal change in the power spectrum naturally occurs for the
largest possible value of $y_{0,max} \equiv k_{max, observed}/a_{0}H$. This is
simply a consequence of the fact that we are studying the effects of
an irrelevant operators whose size increases with $\vk$. 
The existence of an
`earliest time' --- the moment where we can no longer trust the
boundary effective action --- suggests that we choose $a_0 =
k_{max}/M$ (we {\em cannot} choose an $a_0$ smaller than that; we
could choose a larger one). Hence $y_{0,max} = M/H$. For this value of we see
that the change in the power spectrum equals
\begin{eqnarray}
  \label{eq:36}
  \frac{P_{BD}^{dS}+\delta P}{P^{dS}_{BD}}(y_{0,max}) &=& 1+
  \frac{\pi}{4}\frac{\beta H}{M} \left[ i \frac{M^2}{H^2}
  \bar{H}_{3/2}^2(M/{H})+{\rm c.c.}\right] \non
&\simeq& 1+ \beta \sin(2{M}/{H})
\end{eqnarray}
Note: thought the chance in the power spectrum is parametrically $H/M$
as argued before, its maximal change is in fact quite independent of
their values --- if one sets $a_0 = k_{max}/M~\leftrightarrow
y_{0,max}=M/H$. For this value of $y_0$, it becomes linearly dependent on the size of the
irrelevant operator $\beta$. We have shown these results in figure
\ref{fig:2}. The observed window in the CMB is four orders of
magnitude from $y_{max}$ to $10^{-4}y_{0,max}$. Clearly for small
values of $\beta$ and moderately large values of $M/H$ the change in
the power spectrum is far larger than the projected $1\%$ uncertainty
in future measurements. We have a solid case that for
a large enough value of $H/M$ future CMB measurements are sensitive to 
high-energy physics through irrelevant corrections to the initial
conditions.

Moreover, figure \ref{fig:2} clearly shows that the current sensitivity with
which the power spectrum is measured already constrains the
allowed values for $\beta$ and $H/M$ in nature. A coarse extrapolation
from the WMAP results \cite{Bennett:2003bz} indicates that the observed power spectrum is scale invariant
with an accuracy of around $10\%$.\footnote{Actual data show a small
  scale dependence. The power spectrum is inversely proportional to a
  slow roll parameter $P \sim 1/\eps$, which is measured with an
  accuracy of about $10\%$. We are
  extrapolating that error here to a hypothetical pure de Sitter
  phase of inflation.} A value of $\beta \sim 0.2$ and $H/M \sim 0.01$
would already imply a  
$20\%$ change at the upper end of the power spectrum, inconsistent
with the data.
 The point of principle that the power spectrum {\em is} sensitive to
 irrelevant corrections has therefore been established.

Naturally, all other --- measured --- cosmological
quantities will also be affected by the irrelevant boundary operators 
and observability therefore hinges on whether other phenomenological
constraints are mild enough to allow a large enough 
change to the power spectrum. 
In particular, an order of magnitude estimate of the gravitational
backreaction \cite{Porrati:2004gz} argued that such constraints are
quite significant.\footnote{That backreaction effects in this context could be important was also emphasized in 
\cite{Tanaka:2000jw} (see also \cite{Lemoine:2001ar}). Other 
phenomenological constraints on initial state 
modifications have been discussed in \cite{Starobinsky:2002rp}. More formal arguments against the use of 
non-standard initial states can be found in \cite{stanford,Einhorn:2002nu}.} These constraints are not in conflict with our arguments in section
\ref{sec:backr-renorm-arbitr}. As stressed there, this is input into
what the correct initial conditions are, from the observed energy
density driving the inflationary expansion.

A forthcoming article will discuss the computation of the
gravitational backreaction in detail. The resulting perturbative 
bound on the
coefficient $\beta$ of the leading irrelevant boundary operator,
\begin{eqnarray}
\label{eq:260a}
|\beta|^2 &\leq& (12 \pi)^2 \, \left({M_p^2 H_0^2 \over
    M_{string}^4}\right)~,
\end{eqnarray}
plus the constraints from the observed inflationary slow-roll parameters $\eps_{observ},
\eta_{observ}$ 
\begin{eqnarray}
\label{eq:11c}
|\beta|^2 &\leq& 2 \, (6 \pi)^2 \, |\epsilon_{observ}| \, \left( {M_p^2 H_0^2 \over M_{string}^4} \right) \\
\label{eq:12c}
|\beta|^2 &\leq& (6 \pi)^2 \, |\epsilon_{observ}| \, |\eta_{observ}| \, \left( {M_p^2 H_0^2 \over M_{string}^4} \right) 
\end{eqnarray}
entail relatively mild backreaction constraints. For typical but
optimistic values for $H \sim 10^{14}$ GeV, the scale of new physics
$M_{string} \sim 10^{16}$ GeV 
and the reduced
Planck mass $M_p \sim 10^{19}$ GeV they allow  
a significant observational window of
opportunity (see figure \ref{fig:2}). 
The mildness results from the fact that the backreaction
is only significantly affected at {\em second} order in the irrelevant
correction. (This had earlier been argued by Tanaka
\cite{Tanaka:2000jw,Starobinsky:2002rp}. Indeed compared to the order of magnitude
estimate \cite{Porrati:2004gz} the above three equations are
effectively the same with $\beta^2$ substituted for $\beta$.)
The backreaction due to the first order correction, though not zero, is
essentially localized on the boundary and therefore subject to the
substraction prescription utilized to renormalize the theory. The
localization is a consequence of the highly oscillatory nature
of the first order power-spectrum. When integrated all contributions
cancel except on the boundary. The second order effect which remains
and dominates is the `time-averaged' energy stored in the oscillatory behaviour
itself. This grows as the square of the amplitude rather than linear,
and it is this which accounts for the appearance of $|\beta|^2$ rather
than $|\beta|$ in eqs. (\ref{eq:260a})-~(\ref{eq:12c}) above.

The bounds on the coefficient $\beta$ due to the one-loop backreaction
are in fact so mild that they are superseded by the direct sensitivity
of the power spectrum for large $H/M$. 
Combining the various sensitivities in figure
\ref{fig:2}, we see how the 
aforementioned existence of an 'earliest time' and its
concommittant bound on $\beta \leq 0.1$ implies that
backreaction poses {\em no} constraints at all if $H/M$ is large
enough. The bounds on $\beta$ from
backreaction are all weaker than the direct `search' 
upper bound from the
power-spectrum. Hence the search is on.

Whether the future data will be of sufficient accuracy to resolve the
contributions of irrelevant corrections to the initial conditions from
other contributions to scale dependence in the power spectrum is a
different question all together. What these results do show is that
such an investigation should be carried out.

\section{Conclusion and Outlook}
\label{sec:conclusion-outlook}
\setcounter{equation}{0}

The recent successes in CMB measurements exemplified by
\cite{Bennett:2003bz}, 
have made the computation of inflationary density perturbations a
focal point of research. The computation of these density
perturbations suffers from a fundamental deficiency, however, that is
at the same time a wondrous opportunity. The
enormous cosmological redshifts push the energy levels beyond 
the bound of validity of
general relativity, the framework in which these computations are
done. From a field theoretic point of view general relativity can be viewed
as the low energy effective action of a more fundamental consistent
theory of quantum gravity. This effective action has higher order
corrections which when re-included increase its range of
validity. These higher order corrections encode the physics that is
specific to quantum gravity. Hence understanding the way these higher order
corrections affect the computation of inflationary density
perturbations is both {needed} to restore consistency to the
computation, and {provides} an opportunity to witness glimpses of
Planck scale physics in a measurable quantity.
 
However, an action by itself is
not sufficient to extract the physics of quantum fields. One must in
addition specify a set of {\em boundary conditions}. Which boundary
conditions to impose is always a physical question. In 
the Hamiltonian
language boundary conditions correspond to a choice of vacuum
state. In cosmological settings, due to the lack of symmetries 
the correct choice of vacuum, i.e. boundary conditions, is
ambiguous. A number of proposals, though, exist for the correct
state. What we have discussed here, is that this vacuum choice
ambiguity can be framed in terms of the arbitrariness of a
boundary action. This puts the full physics in the form of a naturally 
coherent effective action. Deriving the power spectrum of inflationary density
perturbations within this framework, the lowest order corrections are 
irrelevant boundary operators of order $H/M_{Planck}$. Because we are
able to use the language of effective field theory, not only is the
parametric dependence of the inflationary perturbation spectrum on
high-energy physics known, the coefficients
are also in principle computable from the high-energy sector that has
been integrated out. RG-principles tell us that {\em generically} this
coefficient will be non-zero, except for very special choices of
initial conditions and high energy completions of the low energy
theory. In cosmological spacetimes in 
particular the Lorentz symmetry which forbids the
appearance of such corrections in flat Minkowski space is absent. 
This makes the prediction that we can potentially observe
Planck scale physics in the cosmic sky quite strong, 
or equivalently the absence of these effects would 
constrain 
the possible high energy completions, i.e. string theory.\footnote{A
  recent article examing non-Gaussian correlations in the power
  spectrum resulting from boundary interactions is in full support of 
this
  conclusion \cite{Porrati:2004dm}}

Several earlier investigations have shown that the effects related to
a choice of initial conditions are not the only way in which
high-energy physics can show up in cosmological measurements. Effects
due to a non-vanishing classical expectation value of high-
\cite{Burgess:2002ub} or low-energy \cite{ShiuWasserman} fields, or a
modified dispersion relation 
(see, e.g. \cite{Brandenberger})
can be of the same order. The former two 
should fit into our framework by
the explicit introduction of sources. The latter presumes an all-order
effective action, which is finite and therefore has a specific kinetic
term $\cF(\Box/\Lam)$. The subleading effects in $\Lam$ obviously
change the two-point correlation function and hence the power
spectrum. In RG-terms a specific choice of regulator function
$\cF(\Box/\Lam)$ corresponds to a specific choice of UV-completion of
the theory. The relevant behaviour is universal and independent of
the choice of $\cF(\Box/\Lam)$, but the irrelevant corrections are not, 
of course. 

The introduction of a boundary action to account for the initial
 conditions, and its behaviour under RG-flow including irrelevant
 corrections begs for a comparison with 
 the idea of holography. The
 latter suggests that (gravitational) theories in 
$d$-dimensional de Sitter space have
 a dual formulation as a (Euclidean boundary) conformal field theory 
of dimension $d-1$
 \cite{Strominger:2001pn,Balasubramanian:2001nb}. The cosmological
 implications of this conjectured correspondence underline the
 universality and robustness of predictions for inflationary density
 perturbations precisely because they are related to RG characteristics in
 the dual 
$d-1$ dimensional 
theory \cite{Larsen:2003pf, Larsen:2002et,
   vanderSchaar:2003sz}. These qualitative similarities are striking,
 but there are crucial differences with the approach put forth here. 
Holography interchanges the IR and UV properties of the
dual theories. The UV physics of a three-dimensional Euclidean field 
theory corresponds to 
the IR of the four-dimensional de Sitter gravity and vice versa. 
The holographic screen where the dual field theory lives 
corresponds to a boundary action in the de Sitter 
future. Its precise position defines the UV cut-off in the 
Euclidean field theory that should completely 
describe the infinite interior (i.e. the past) of the de Sitter 
bulk gravity theory. Time evolution in the
bulk is then interpreted as RG-flow in the boundary field theory, 
and so the IR physics in the field theory 
corresponds to the infinite past in the bulk. Instead the boundary 
actions considered in this paper are introduced only to encode 
the initial conditions in the past of the four 
dimensional de Sitter gravity theory. They are not dual 
descriptions of the bulk de Sitter theory, but are merely introduced 
as effective tools to describe the initial conditions in the bulk.  
Nevertheless, it would be very interesting to study how 
the results described in this paper should be interpreted from the point
of view of a putative dual three-dimensional Euclidean field theory.

The boundary effective action encoding the initial conditions finally answers
the longstanding open question: do cut-off theories in a cosmological
setting cease to be valid beyond an earliest time? Naively this is
so. The results here show that the blueshifted momentum expansion on the
boundary effective action supplies the mathematical underpinning for
this intuition. This time, though clearly a fiducial one, is a natural
location for our boundary action. 
The freedom, however, remains to impose initial conditions where-ever
one wishes. We may have chosen any other fiducial point as long as the
momentum expansion stays under control. What is clear is that 
the choice of this point is immaterial to the issue of boundary conditions in FRW
universes. This fact is made manifest in the symmetry
(\ref{eq:14}) between boundary location $y_0$ and boundary coupling
$\kap$. Physics depends only on the invariant combination $b_{\kap}(y_0)$. 
With the effective field theory
description in mind, and the idea that `vacua' are boundary RG fixed
points, a truly interesting question is whether such boundary
conditions exist, and if so, how they are related to the known cosmological
vacuum choices.

\subsection{A comparison with previous results and the discussion 
on $\alpha$-states}

Much discussion has taken place in the recent literature on the
consistency of 
so-called $\alp$-states in de Sitter space
\cite{stanford, Einhorn:2002nu}. Initial investigations into the sensitivity of
inflationary perturbations to high energy physics found that in pure de Sitter
the leading $H/M$ corrections to the power spectrum 
can be interpreted as choosing the harmonic oscillator vacuum (section
\ref{sec:bound-cond-adiab}) at the naive earliest time 
$
\eta_0(\vk) = -\Lam/H | \vk |$ 
where the theory makes sense, rather than the
Bunch-Davies choice
\cite{egks, Danielsson:2002kx}. 
Imposing such boundary conditions in pure de Sitter can equivalently 
be interpreted as selecting a non-trivial de Sitter invariant vacuum 
state called an $\alpha$-state \cite{Danielsson:2002kx}. Strictly
speaking, the Shortest Length (SL) boundary conditions are only imposed
on momentum modes below the cut-off scale $\Lam$ of the theory, and
they are not true de Sitter $\alp$-states. 
Subject 
to this distinction, 
the purported inconsistency of
$\alp$-states, {particularly with respect to the decoupling of Planck scale physics}
\cite{Einhorn:2002nu},
therefore would have major consequences (see, however,
\cite{deBoer:2004nd}). If $\alp$-states and 
other boundary
conditions are all
inconsistent, all high-energy
physics would have to be encoded in bulk irrelevant operators. This
would put
transplanckian effects in the CMB perturbation spectrum beyond
observational reach. 

Let us put first, that our results form solid evidence for the
presence of $H/M$ effects affecting inflationary predictions for the
CMB perturbation
spectrum. As the explicit expression (\ref{eq:75}) we derive for the
power spectrum 
shows, our results, though qualitatively similar, are quantitatively
far more general 
from having `chosen' an (cut-off) $\alp$-state. 
The
coherent 
effective Lagrangian approach followed here
gives a precise answer which differs in general from the
(earliest-time) $\alp$-state approach,
but upholds the qualitative validity. One can certainly ask to what
choice of `vacuum state' our results correspond; given the physical
parameter $b_{\kap}$ this is straightforward to work out. The answer may
be interesting from the point of view of Hamiltonian dynamics, but 
as we have shown here, in the Lagrangian language of boundary
conditions, any initial state which can be described by a local
relevant boundary
coupling $\kap$ is consistent.
{\em There is no need to know whether
  $\alp$-states are consistent to study transplanckian corrections to
  inflationary perturbations.}

At the same time, vacuum choices, $\alp$-states included, do
correspond to boundary conditions.\footnote{We are grateful to Brian
  Greene both for emphasizing the importance in explicitly discussing
  the consistency of $\alp$-vacua and his help in resolving the
  issue.} 
 And boundary conditions should not 
 spoil decoupling, although there will be new effects, as we reviewed in
section \ref{sec:deco-theor-with}. Taking this lesson to heart, 
it is hard to see how (earliest-time) $\alpha$-states could be inconsistent.  
A recent article \cite{Collins:2003mj} arguing for the
consistency of $\alpha$-vacua does not exactly follow 
the approach outlined here, but is very much in the spirit of
introducing  boundary
counterterms. 
An answer, however, is provided by pursuing the discussion in section
\ref{sec:bound-cond-adiab} further. 
The (cut-off) $\alp$-vacua correspond to
choosing earliest-time boundary conditions in an effective theory below
scale $M$ with the physical parameter $b_{SL}$ a constant
number. The precise relation is that $b_{SL}=e^{\alp}$. 
One then readily derives that an $\alp$-vacuum corresponds to
a boundary coupling (see eq. (\ref{eq:85}))
\begin{eqnarray}
  \label{eq:88}
  \kap_{SL} =
  -\frac{\pa_n\phi_+(\eta'_0)+b_{SL}\pa_n\phi_-(\eta'_0)}{\phi_+(\eta'_0)+b_{SL}\phi_-(\eta'_0)} ~.
\end{eqnarray}
Recall that $b_{SL}$ is constant. To analyze the high spatial momentum
behavior, we may therefore approximate the modefunctions
$\phi_{\pm}(\eta_0')$ by their Minkowski counterparts.
In this limit 
the
boundary coupling $\kap_{SL}$ encoding $\alp$-states becomes
\begin{eqnarray}
  \label{eq:102}
|\vk| \rar \infty,~~~ \kap_{SL} &\simeq&
 -i\frac{|\vk|}{a_0}\frac{e^{i|\vk|\eta_0'}-b_{SL}e^{-i|\vk|\eta_0'}}{e^{i|\vk|\eta_0'}
+b_{SL}e^{-i|\vk|\eta_0'}}  ~.
\end{eqnarray}
The boundary coupling $\kap_{SL}$ therefore has an infinite set of 
poles 
\begin{equation}
|\vk|=\frac{-1}{2\eta_0'}\left((2n+1)\pi+i\ln(b_{SL})\right) \, , 
~~~~
n\in \ZZ~,
\end{equation}
in the momentum plane. Clearly 
this boundary coupling corresponds to a non-local action. Cut-off
$\alp$-states, i.e. shortest length boundary conditions, therefore 
fall outside the class of local relevant boundary conditions 
we study here.
But are they
inconsistent? Recall that the original studies
\cite{egks,Danielsson:2002kx} argue that $\alp$-vacua
should encode (first order) effects of high-energy physics in the
spectrum of inflationary density perturbations. 
This
point of view therefore states that by construction the boundary
coupling 
$\kap_{SL}$ includes the effects of irrelevant {\em boundary
  operators}. We are therefore instructed to treat the non-local
nature of the boundary coupling $\kap_{SL}$ in
the low-energy effective action in
the usual way. One expands around the origin $|\vk|=0$ in the momentum
plane generating a series of higher derivative 
irrelevant boundary operators with specific leading
coefficients $\bet_i$.\footnote{It is not completely clear that this
  interpretation withstands close scrutiny. Most
  non-local terms in the effective action have real poles. Here we
  are confronted with imaginary poles. Perhaps $\alp$-vacua correspond
  to a high-energy completion with numerous unstable particles.} 
This expansion is valid as long as we limit the
range of our effective action to the location of the first pole $|\vk|
=\frac{1}{2|\eta_0'|}\sqrt{|\pi+i\ln b_{SL}|^2}$, i.e. physical momenta
are constrained to the range $|p_0| =|\frac{\vk}{a_0}| \lesssim
\frac{H}{2}|\ln b_{SL}|$. (Eq. (\ref{eq:12}) gives us 
$b_{SL}\simeq
  H/2Me^{-2iM/H-i\pi/2}$, and we recover the cut-off $|p|<M$.)
The fact that the complicated pole structure of boundary
couplings of alpha-vacua
is highly specific (they ensure that (non-cut-off) 
$\alp$-vacua are invariant under
de Sitter isometries) is not to the point in this perspective.  
It is then also clear why $\alp$-vacua are not
renormalizable, in particular in the sense that the bare 
backreaction, the
divergence in the stress tensor, is to leading order not identical to
that in Minkowski space. Irrelevant operators correspond to
non-renormalizable terms in the action. Because the pole structure of
the boundary coupling $\kap$ reveals that $\alp$-states are correctly 
to be interpreted as encoding specific contributions from irrelevant
operators, any correlation function computed with respect to the
$\alp$-vacuum, includes the contribution from these irrelevant
operators. It is therefore {\em expected} to be
non-renormalizable. Obviously this does not mean that the $\alp$-vacua
are inconsistent. As always in effective actions 
one
must `neglect' any contributions of irrelevant operators 
for the purposes of renormalization. They only
make sense in a theory with a manifest cut-off
\cite{Polchinski}. Removing the cut-off, removes the irrelevant
operators. Indeed the $\alp$-states proposed in
\cite{egks,Danielsson:2002kx} with $b_{SL} \simeq H/2M$ are
naturally in accordance with this precept.

In this sense, the (cut-off) 
$\alp$-vacua are therefore manifestly consistent in
the framework put forth 
here. They simply correspond 
to a specific choice of leading and higher irrelevant
boundary operators. Whatever they are is not very interesting from the
perspective of effective field theory.\footnote{They are
  $\beta_c=-\frac{i}{3} e^{-2iM/H},\bpa-\bpe=\frac{7i}{3}e^{-2iM/H}$. 
Expanding around small $|b_{SL}|=H/2M$ and small $ |\vk| \ll Ha_0$, we see that
\begin{eqnarray}
\nonumber
\kap_{SL} &=&
-i\frac{\vk}{a_0}\left[1-\frac{H}{iM}e^{-2iM/H}  - \frac{2
    |\vk|}{a_0M} e^{-2iM/H}\right] 
=
\kap_{BD} - \kap_{BD}\frac{H}{iM}e^{-2i M/H}  -2i
\frac{\kap^2_{BD}}{M}e^{-2iM/H} 
\end{eqnarray}
Comparing with~(\ref{eq:5}) we find the coefficients $\beta_i$.} 
A specific choice for the irrelevant operators means having chosen a
specific form for the high-energy transplanckian completion of the
theory. But what this physics is, is precisely the knowledge we are
after. 

\begin{theacknowledgments}
We thank Robert Brandenberger, Cliff Burgess, Chong-Sun Chu, Jim Cline, 
Richard Easther, Brian Greene and Erick Weinberg for comments. We are 
particularly grateful to the organizers and participants of the
Amsterdam Summer Workshops on String Theory 2002 and 2003 and those of
the String Cosmology Conference at Santa Barbara, October 2003. KS and JPvdS
both thank the string theory group at the University of Wisconsin 
at Madison for hospitality. Likewise, GS thanks ISCAP and the theory 
group at Columbia University. KS acknowledges financial support from 
DOE grant DE-FG-02-92ER40699. The work of GS was supported in part by 
NSF CAREER Award No. PHY-0348093, a Research Innovation Award from Research 
Corporation and in part by funds from the University of Wisconsin.
\end{theacknowledgments}


{\small
\begin{thebibliography}{99}

\bibitem{Brandenberger}
R.~H.~Brandenberger,
{\em Inflationary cosmology: Progress and problems},
arXiv:hep-ph/9910410.
%%CITATION = HEP-PH 9910410;%%
\\
J.~Martin and R.~H.~Brandenberger,
{\em The trans-Planckian problem of inflationary cosmology},
Phys.\ Rev.\ D {\bf 63}, 123501 (2001)
[arXiv:hep-th/0005209].
%%CITATION = HEP-TH 0005209;%%
\\
R.~H.~Brandenberger and J.~Martin,
{\em The robustness of inflation to changes in super-Planck-scale physics},
Mod.\ Phys.\ Lett.\ A {\bf 16}, 999 (2001)
[arXiv:astro-ph/0005432].
%%CITATION = ASTRO-PH 0005432;%%
\\
J.~C.~Niemeyer,
{\em Inflation with a high frequency cutoff}
Phys.\ Rev.\ D {\bf 63}, 123502 (2001)
[arXiv:astro-ph/0005533].
%%CITATION = ASTRO-PH 0005533;%%

\bibitem{egks}
R.~Easther, B.~R.~Greene, W.~H.~Kinney and G.~Shiu,
{\em Inflation as a probe of short distance physics},
Phys.\ Rev.\ D {\bf 64}, 103502 (2001)
[arXiv:hep-th/0104102].
%%CITATION = HEP-TH 0104102;%%
\\
R.~Easther, B.~R.~Greene, W.~H.~Kinney and G.~Shiu,
{\em Imprints of short distance physics on inflationary cosmology},
arXiv:hep-th/0110226.
%%CITATION = HEP-TH 0110226;%%
\\
R.~Easther, B.~R.~Greene, W.~H.~Kinney and G.~Shiu,
{\em A generic estimate of trans-Planckian modifications to the
  primordial  power spectrum in inflation}, 
Phys.\ Rev.\ D {\bf 66}, 023518 (2002)
[arXiv:hep-th/0204129].
%%CITATION = HEP-TH 0204129;%%
\\
C.~S.~Chu, B.~R.~Greene and G.~Shiu,
{\em Remarks on inflation and noncommutative geometry},
Mod.\ Phys.\ Lett.\ A {\bf 16}, 2231 (2001)
[arXiv:hep-th/0011241].
%%CITATION = HEP-TH 0011241;%%

\bibitem{ShiuWasserman}
G.~Shiu and I.~Wasserman,
{\em On the signature of short distance scale in the cosmic microwave
background},
Phys.\ Lett.\ B {\bf 536}, 1 (2002)
[arXiv:hep-th/0203113].
%%CITATION = HEP-TH 0203113;%%

\bibitem{stanford}
N.~Kaloper, M.~Kleban, A.~E.~Lawrence and S.~Shenker,
{\em Signatures of short distance physics in the cosmic microwave
  background}, 
Phys.\ Rev.\ D {\bf 66}, 123510 (2002)
[arXiv:hep-th/0201158].
%%CITATION = HEP-TH 0201158;%%
\\
N.~Kaloper, M.~Kleban, A.~Lawrence, S.~Shenker and L.~Susskind,
{\em Initial conditions for inflation},
JHEP {\bf 0211}, 037 (2002)
[arXiv:hep-th/0209231].
%%CITATION = HEP-TH 0209231;%%

\bibitem{Others}
J.~Martin and R.~H.~Brandenberger,
{\em A cosmological window on trans-Planckian physics},
arXiv:astro-ph/0012031.
%%CITATION = ASTRO-PH 0012031;%%
\\
A.~Kempf,
{\em Mode generating mechanism in inflation with cutoff},
Phys.\ Rev.\ D {\bf 63}, 083514 (2001)
[arXiv:astro-ph/0009209].
\\
J.~C.~Niemeyer and R.~Parentani,
{\em Trans-Planckian dispersion and scale-invariance of inflationary
  perturbations},
Phys.\ Rev.\ D {\bf 64}, 101301 (2001)
[arXiv:astro-ph/0101451].
%%CITATION = ASTRO-PH 0101451;%%
\\
A.~Kempf and J.~C.~Niemeyer,
{\em Perturbation spectrum in inflation with cutoff},
Phys.\ Rev.\ D {\bf 64}, 103501 (2001)
[arXiv:astro-ph/0103225].
%%CITATION = ASTRO-PH 0103225;%%
\\
M.~Bastero-Gil,
{\em What can we learn by probing trans-Planckian physics},
arXiv:hep-ph/0106133.
%%CITATION = HEP-PH 0106133;%%
\\
L.~Hui and W.~H.~Kinney,
{\em Short distance physics and the consistency relation for scalar and  tensor
fluctuations in the inflationary universe},
Phys.\ Rev.\ D {\bf 65}, 103507 (2002)
[arXiv:astro-ph/0109107].
%%CITATION = ASTRO-PH 0109107;%%
\\
R.~H.~Brandenberger, S.~E.~Joras and J.~Martin,
{\em Trans-Planckian physics and the spectrum of fluctuations in a bouncing  universe},
Phys.\ Rev.\ D {\bf 66}, 083514 (2002)
[arXiv:hep-th/0112122].
%%CITATION = HEP-TH 0112122;%%
\\
J.~Martin and R.~H.~Brandenberger,
{\em The Corley-Jacobson dispersion relation and trans-Planckian inflation},
Phys.\ Rev.\ D {\bf 65}, 103514 (2002)
[arXiv:hep-th/0201189].
%%CITATION = HEP-TH 0201189;%%
\\
R.~Brandenberger and P.~M.~Ho,
{\em Noncommutative spacetime, stringy spacetime uncertainty principle, and
 density fluctuations},
Phys.\ Rev.\ D {\bf 66}, 023517 (2002)
[AAPPS Bull.\  {\bf 12N1}, 10 (2002)]
[arXiv:hep-th/0203119].
%%CITATION = HEP-TH 0203119;%%
\\
S.~Shankaranarayanan,
{\em Is there an imprint of Planck scale physics on inflationary cosmology?},
Class.\ Quant.\ Grav.\  {\bf 20}, 75 (2003)
[arXiv:gr-qc/0203060].
%%CITATION = GR-QC 0203060;%%
\\
S.~F.~Hassan and M.~S.~Sloth,
{\em Trans-Planckian effects in inflationary cosmology and the modified  uncertainty principle},
Nucl.\ Phys.\ B {\bf 674}, 434 (2003)
[arXiv:hep-th/0204110].
%%CITATION = HEP-TH 0204110;%%
\\
J.~C.~Niemeyer, R.~Parentani and D.~Campo,
{\em Minimal modifications of the primordial power spectrum from an
  adiabatic short distance cutoff},
Phys.\ Rev.\ D {\bf 66}, 083510 (2002)
[arXiv:hep-th/0206149].
%%CITATION = HEP-TH 0206149;%%
\\
K.~Goldstein and D.~A.~Lowe,
{\em Initial state effects on the cosmic microwave background and
  trans-planckian physics},
Phys.\ Rev.\ D {\bf 67}, 063502 (2003)
[arXiv:hep-th/0208167].
%%CITATION = HEP-TH 0208167;%%
\\
V.~Bozza, M.~Giovannini and G.~Veneziano,
{\em Cosmological perturbations from a new-physics hypersurface},
JCAP {\bf 0305}, 001 (2003)
[arXiv:hep-th/0302184].
%%CITATION = HEP-TH 0302184;%%
\\
G.~L.~Alberghi, R.~Casadio and A.~Tronconi,
{\em Trans-Planckian footprints in inflationary cosmology},
Phys.\ Lett.\ B {\bf 579}, 1 (2004)
[arXiv:gr-qc/0303035].
%%CITATION = GR-QC 0303035;%%
\\
J.~Martin and R.~Brandenberger,
{\em On the dependence of the spectra of fluctuations in inflationary
  cosmology on trans-Planckian physics},
Phys.\ Rev.\ D {\bf 68}, 063513 (2003)
[arXiv:hep-th/0305161].
%%CITATION = HEP-TH 0305161;%%
\\
N.~Kaloper and M.~Kaplinghat,
{\em Primeval corrections to the CMB anisotropies},
arXiv:hep-th/0307016.
%%CITATION = HEP-TH 0307016;%%

\bibitem{Tanaka:2000jw}
T.~Tanaka,
{\em A comment on trans-Planckian physics in inflationary universe},
arXiv:astro-ph/0012431.
%%CITATION = ASTRO-PH 0012431;%%

\bibitem{Lemoine:2001ar}
M.~Lemoine, M.~Lubo, J.~Martin and J.~P.~Uzan,
{\em The stress-energy tensor for trans-Planckian cosmology},
Phys.\ Rev.\ D {\bf 65} (2002) 023510
[arXiv:hep-th/0109128].
%%CITATION = HEP-TH 0109128;%%

\bibitem{Starobinsky:2002rp}
A.~A.~Starobinsky,
{\em Robustness of the inflationary perturbation spectrum to
  trans-Planckian physics},
Pisma Zh.\ Eksp.\ Teor.\ Fiz.\  {\bf 73}, 415 (2001)
[JETP Lett.\  {\bf 73}, 371 (2001)]
[arXiv:astro-ph/0104043].
%%CITATION = ASTRO-PH 0104043;%%
\\
A.~A.~Starobinsky and I.~I.~Tkachev,
{\em Trans-Planckian particle creation in cosmology and ultra-high
  energy cosmic rays},
JETP Lett.\  {\bf 76}, 235 (2002)
[Pisma Zh.\ Eksp.\ Teor.\ Fiz.\  {\bf 76}, 291 (2002)]
[arXiv:astro-ph/0207572].
%%CITATION = ASTRO-PH 0207572;%%



\bibitem{Danielsson:2002kx}
U.~H.~Danielsson,
{\em A note on inflation and transplanckian physics},
Phys.\ Rev.\ D {\bf 66}, 023511 (2002)
[arXiv:hep-th/0203198]. 
%%CITATION = HEP-TH 0203198;%%
\\
U.~H.~Danielsson,
{\em Inflation, holography and the choice of vacuum in de Sitter space},
JHEP {\bf 0207}, 040 (2002)
[arXiv:hep-th/0205227].
%%CITATION = HEP-TH 0205227;%%

\bibitem{Burgess:2002ub}
C.~P.~Burgess, J.~M.~Cline, F.~Lemieux and R.~Holman,
{\em Are inflationary predictions sensitive to very high energy physics?},
JHEP {\bf 0302}, 048 (2003)
[arXiv:hep-th/0210233].
%%CITATION = HEP-TH 0210233;%%
\\
C.~P.~Burgess, J.~M.~Cline and R.~Holman,
{\em Effective field theories and inflation},
arXiv:hep-th/0306079.
%%CITATION = HEP-TH 0306079;%%
\\
C.~P.~Burgess, J.~M.~Cline, F.~Lemieux and R.~Holman,
{\em Decoupling, trans-Planckia and inflation},
arXiv:astro-ph/0306236.
%%CITATION = ASTRO-PH 0306236;%%

\bibitem{Bennett:2003bz}
C.~L.~Bennett {\it et al.},
{\em First Year Wilkinson Microwave Anisotropy Probe (WMAP)
  Observations: Preliminary Maps and Basic Results}, 
Astrophys.\ J.\ Suppl.\  {\bf 148}, 1 (2003)
[arXiv:astro-ph/0302207].
%%CITATION = ASTRO-PH 0302207;%%
\\
G.~Hinshaw {\it et al.},
{\em First Year Wilkinson Microwave Anisotropy Probe (WMAP)
  Observations: Angular Power Spectrum}, 
Astrophys.\ J.\ Suppl.\  {\bf 148}, 135 (2003)
[arXiv:astro-ph/0302217].
%%CITATION = ASTRO-PH 0302217;%%
\\
H.~V.~Peiris {\it et al.},
{\em First year Wilkinson Microwave Anisotropy Probe (WMAP)
  observations:  Implications for inflation}, 
Astrophys.\ J.\ Suppl.\  {\bf 148}, 213 (2003)
[arXiv:astro-ph/0302225].
%%CITATION = ASTRO-PH 0302225;%%

\bibitem{nous}
K.~Schalm, G.~Shiu and J.~P.~van der Schaar,
{\em Decoupling in an expanding universe: Boundary RG-flow affects initial
conditions for inflation},
JHEP {\bf 0404}, 076 (2004)
[arXiv:hep-th/0401164].
%%CITATION = HEP-TH 0401164;%%

\bibitem{BD}
N.~D.~Birrell and P.~C.~W.~Davies,
{\em Quantum Fields In Curved Space}, Cambridge University Press, 1982.
%\href{http://www.slac.stanford.edu/spires/find/hep/www?irn=998621}{SPIRES entry}

\bibitem{Larsen:2003pf}
F.~Larsen and R.~McNees,
{\em Inflation and de Sitter holography},
arXiv:hep-th/0307026.
%%CITATION = HEP-TH 0307026;%%

\bibitem{Burgess2}
C.~P.~Burgess,
{\em Quantum gravity in everyday life: General relativity as an effective field theory},
arXiv:gr-qc/0311082.
%%CITATION = GR-QC 0311082;%%

\bibitem{Brandenberger:2002hs}
R.~H.~Brandenberger and J.~Martin,
{\em On signatures of short distance physics in the cosmic microwave
background},
Int.\ J.\ Mod.\ Phys.\ A {\bf 17}, 3663 (2002)
[arXiv:hep-th/0202142].
%%CITATION = HEP-TH 0202142;%%

\bibitem{Symanzik:1981wd}
K.~Symanzik,
{\em Schrodinger Representation And Casimir Effect In Renormalizable Quantum Field Theory,}
Nucl.\ Phys.\ B {\bf 190}, 1 (1981).
%%CITATION = NUPHA,B190,1;%%

\bibitem{Candelas:qw}
P.~Candelas,
{\em Vacuum Polarization In The Presence Of Dielectric And Conducting
  Surfaces},
Annals Phys.\  {\bf 143}, 241 (1982).
%%CITATION = APNYA,143,241;%%

\bibitem{Deutsch:sc}
D.~Deutsch and P.~Candelas,
{\em Boundary Effects In Quantum Field Theory},
Phys.\ Rev.\ D {\bf 20}, 3063 (1979).
%%CITATION = PHRVA,D20,3063;%%

\bibitem{Barton:1979yd}
G.~Barton,
{\em Some Surface Effects In The Hydrodynamic Model Of Metals},
Rept.\ Prog.\ Phys.\  {\bf 42}, 963 (1979).
%%CITATION = RPPHA,42,963;%%

\bibitem{Witten:2001ua}
E.~Witten,
{\em Multi-trace operators, boundary conditions, and AdS/CFT
  correspondence},
arXiv:hep-th/0112258.
%%CITATION = HEP-TH 0112258;%%

\bibitem{Fredenhagen:2003xf}
S.~Fredenhagen,
{\em Organizing boundary RG flows},
Nucl.\ Phys.\ B {\bf 660}, 436 (2003)
[arXiv:hep-th/0301229].
%%CITATION = HEP-TH 0301229;%%

\bibitem{Aharony:2003qf}
O.~Aharony, O.~DeWolfe, D.~Z.~Freedman and A.~Karch,
{\em Defect conformal field theory and locally localized gravity},
JHEP {\bf 0307}, 030 (2003)
[arXiv:hep-th/0303249].
%%CITATION = HEP-TH 0303249;%%

\bibitem{Burgess1}
Y.~Aghababaie and C.~P.~Burgess,
{\em Effective actions, boundaries and precision calculations of Casimir energies},
arXiv:hep-th/0304066.
%%CITATION = HEP-TH 0304066;%%

\bibitem{Graham:2003nc}
K.~Graham and G.~M.~T.~Watts,
{\em Defect lines and boundary flows},
arXiv:hep-th/0306167.
%%CITATION = HEP-TH 0306167;%%

\bibitem{Jaffe:2003ji}
R.~L.~Jaffe,
{\em Unnatural acts: Unphysical consequences of imposing boundary
  conditions  on quantum fields}, 
AIP Conf.\ Proc.\  {\bf 687}, 3 (2003)
[arXiv:hep-th/0307014].
%%CITATION = HEP-TH 0307014;%%

\bibitem{Leigh:jq}
R.~G.~Leigh,
{\em Dirac-Born-Infeld Action From Dirichlet Sigma Model},
Mod.\ Phys.\ Lett.\ A {\bf 4}, 2767 (1989).
%%CITATION = MPLAE,A4,2767;%%

\bibitem{Polchinski}
J.~Polchinski,
{\em Renormalization And Effective Lagrangians},
Nucl.\ Phys.\ B {\bf 231}, 269 (1984).
%%CITATION = NUPHA,B231,269;%%

\bibitem{Hamilton:2003xr}
A.~J.~Hamilton, D.~Kabat and M.~K.~Parikh,
{\em Cosmological particle production without Bogolubov coefficients},
arXiv:hep-th/0311180.
%%CITATION = HEP-TH 0311180;%%

\bibitem{Chung:2003wn}
D.~J.~H.~Chung, A.~Notari and A.~Riotto,
{\em Minimal theoretical uncertainties in inflationary predictions},
JCAP {\bf 0310}, 012 (2003)
[arXiv:hep-ph/0305074].
%%CITATION = HEP-PH 0305074;%

\bibitem{Porrati:2004gz}
M.~Porrati,
{\em Bounds on generic high-energy physics modifications to the
  primordial power spectrum from back-reaction on the metric}, 
arXiv:hep-th/0402038.
%%CITATION = HEP-TH 0402038;%%

\bibitem{Einhorn:2002nu}
T.~Banks and L.~Mannelli,
{\em De Sitter vacua, renormalization and locality},
Phys.\ Rev.\ D {\bf 67}, 065009 (2003)
[arXiv:hep-th/0209113].
%%CITATION = HEP-TH 0209113;%%
\\
M.~B.~Einhorn and F.~Larsen,
{\em Interacting quantum field theory in de Sitter vacua},
Phys.\ Rev.\ D {\bf 67}, 024001 (2003)
[arXiv:hep-th/0209159].
%%CITATION = HEP-TH 0209159;%%
\\
K.~Goldstein and D.~A.~Lowe,
{\em A note on alpha-vacua and interacting field theory in de Sitter space},
Nucl.\ Phys.\ B {\bf 669}, 325 (2003)
[arXiv:hep-th/0302050].
%%CITATION = HEP-TH 0302050;%%
\\
M.~B.~Einhorn and F.~Larsen,
{\em Squeezed states in the de Sitter vacuum},
Phys.\ Rev.\ D {\bf 68}, 064002 (2003)
[arXiv:hep-th/0305056].
%%CITATION = HEP-TH 0305056;%%
\\
H.~Collins, R.~Holman and M.~R.~Martin,
{\em The fate of the alpha-vacuum},
arXiv:hep-th/0306028.
%%CITATION = HEP-TH 0306028;%%
\\
K.~Goldstein and D.~A.~Lowe,
{\em Real-time perturbation theory in de Sitter space},
arXiv:hep-th/0308135.
%%CITATION = HEP-TH 0308135;%%
\\
H.~Collins and M.~R.~Martin,
{\em The enhancement of inflaton loops in an alpha-vacuum},
arXiv:hep-ph/0309265.
%%CITATION = HEP-PH 0309265;%%


\bibitem{Porrati:2004dm}
M.~Porrati,
{\em Effective field theory approach to cosmological initial
  conditions: Self-consistency bounds and non-Gaussianities},
arXiv:hep-th/0409210.
%%CITATION = HEP-TH 0409210;%%


\bibitem{Strominger:2001pn}
A.~Strominger,
{\em The dS/CFT correspondence},
JHEP {\bf 0110}, 034 (2001)
[arXiv:hep-th/0106113].
%%CITATION = HEP-TH 0106113;%%

\bibitem{Balasubramanian:2001nb}
V.~Balasubramanian, J.~de Boer and D.~Minic,
{\em Mass, entropy and holography in asymptotically de Sitter spaces},
Phys.\ Rev.\ D {\bf 65}, 123508 (2002)
[arXiv:hep-th/0110108].
%%CITATION = HEP-TH 0110108;%%

\bibitem{Larsen:2002et}
F.~Larsen, J.~P.~van der Schaar and R.~G.~Leigh,
{\em de Sitter holography and the cosmic microwave background},
JHEP {\bf 0204}, 047 (2002)
[arXiv:hep-th/0202127].
%%CITATION = HEP-TH 0202127;%%

\bibitem{vanderSchaar:2003sz}
J.~P.~van der Schaar,
{\em Inflationary perturbations from deformed CFT},
arXiv:hep-th/0307271.
%%CITATION = HEP-TH 0307271;%%

\bibitem{deBoer:2004nd}
J.~de Boer, V.~Jejjala and D.~Minic,
{\em Alpha-states in de Sitter space},
arXiv:hep-th/0406217.
%%CITATION = HEP-TH 0406217;%%

%\bibitem{Lidsey}
%J.~E.~Lidsey, A.~R.~Liddle, E.~W.~Kolb, E.~J.~Copeland, T.~Barreiro
%and M.~Abney, 
%{\em Reconstructing the inflaton potential: An overview},
%Rev.\ Mod.\ Phys.\  {\bf 69}, 373 (1997)
%[arXiv:astro-ph/9508078].
%%CITATION = ASTRO-PH 9508078;%%

\bibitem{Collins:2003mj}
H.~Collins and R.~Holman,
{\em Taming the alpha-vacuum},
arXiv:hep-th/0312143.
%%CITATION = HEP-TH 0312143;%%

\bibitem{br}
B.~R.~Greene, K.~Schalm, G.~Shiu and J.~P.~van der Schaar,
{\em Decoupling in an expanding universe: Backreaction barely
  constrains short distance effects in the CMB},
arXiv:hep-th/0411217.
%%CITATION = HEP-TH 0411217;%%


\end{thebibliography}         
}
\end{document}